\documentclass[11pt,a4paper]{article}
\usepackage[margin=25mm]{geometry}
\usepackage[T1]{fontenc}
\usepackage{lmodern,microtype}
\usepackage{amsmath,amssymb,bm,booktabs,graphicx}
\usepackage{float,algpseudocode,longtable}
\algrenewcommand\algorithmicrequire{\textbf{Input:}}
\usepackage{authblk}
\floatstyle{ruled}
\newfloat{algorithm}{tbp}{loa}
\floatname{algorithm}{Algorithm}
\usepackage[round,authoryear]{natbib}
\usepackage[hidelinks]{hyperref}
\usepackage{placeins}
\makeatletter
\newcommand{\pairedtablecaption}{\def\@captype{table}\caption}
\makeatother
\usepackage{comment}
\newcommand{\E}{\mathbb{E}}

\newcommand{\logitinv}{\operatorname{logit}^{-1}}
\newcommand{\RMSE}{\operatorname{RMSE}}
\newcommand{\RMSSE}{\operatorname{RMSSE}}

\title{Joint Estimation of Abilities and Item Parameters for Large Language Models Using Probability and Binary Responses:\\A Fractional-Response IRT Framework}
\author[1]{Masayuki Takayama\thanks{Corresponding author: \texttt{masayuki-takayama@biwako.shiga-u.ac.jp}.}}
\author[1,2,3]{Shohei Shimizu}

\affil[1]{Data Science and AI Innovation Research Promotion Center, Shiga University, Japan}
\affil[2]{SANKEN, The University of Osaka, Japan}
\affil[3]{Faculty of Data Science, Shiga University, Japan}
\date{}
\begin{document}
\maketitle

\begin{abstract}
Large language model (LLM) evaluations provide two response types: probabilities assigned to correct answers and binary indicators of correctness. 
Restricting evaluation to one type narrows the range of available models or discards probability information. 
We propose a method to integrate both types through a common two-parameter logistic item response model and fractional-response quasi-maximum likelihood estimation. 
The framework includes binary-only models in common-scale evaluation without discarding available probability information; broader measurement coverage does not by itself ensure greater precision or measurement equivalence. 
Simulations evaluate statistical efficiency when both response types have the same expected value. 
Moreover, adding probability respondents improves item calibration more efficiently than adding binary respondents. 
However, adding noisier binary responses to an already precise probability sample can increase estimation error under equal observation weights.
%Ability estimates for the original respondents change little in either direction. 
An application to Massive Multitask Language Understanding (MMLU) combines probability response data from the Open LLM Leaderboard with binary response data from Holistic Evaluation of Language Models (HELM) and reveals model-family response patterns, differences in item characteristics, and systematic ability shifts between evaluation protocols.
%Analytical standard errors (SEs) and bootstrap checks assess uncertainty.
The proposed framework extends common-scale IRT evaluation to models with different response formats while highlighting the roles of observation variability and evaluation-protocol comparability in statistical efficiency and interpretation.

\end{abstract}

\section{Introduction}

\paragraph{Background}\mbox{}\

Large language models (LLMs) are increasingly evaluated across a wide range of academic and applied domains. 
In particular, multiple-choice benchmarks such as Massive Multitask Language Understanding (MMLU)~\citep{hendrycks2021} provide a convenient way to compare models across many subject areas. 
Such evaluations are commonly summarized by average accuracy. 
Accuracy is useful as an aggregate performance measure, but it treats all questions equally and does not distinguish between questions that are easy or difficult for nearly all models and questions that sharply differentiate models. 
In this respect, item response theory (IRT) provides a complementary measurement framework by jointly characterizing respondent ability\footnote{
In this paper, ``ability'' is a latent measure of performance within a specified benchmark domain and evaluation protocol; it should not be interpreted as a general measure of intelligence or reasoning competence.
}
and item properties such as difficulty and discrimination. 

Applying IRT to LLM evaluation, however, differs from conventional human testing in an important way. 
The number of distinct LLM respondents available for a particular evaluation can be limited relative to the number of benchmark items, and the precision of item calibration depends primarily on the respondent sample. 
Recent simulation evidence identifies this imbalance, together with non-normal model-ability distributions, as a challenge for reliable IRT-based AI evaluation \citep{jiang2026}.
Repeatedly querying the same fixed LLM, without updating its model parameters,\footnote{
In conventional human testing, obtaining repeated responses to the same item under approximately independent conditions is difficult because previous exposure may affect subsequent responses. 
For a fixed LLM, repeated queries can instead be issued in separate sessions under the same prompts and rules for selecting or sampling the output, making approximately independent repeated measurements easier to design, although independence is not guaranteed automatically.
Reported log probabilities need not be deterministic: \citet{andriushchenko2025} report variation across identical GPT-4 Turbo queries even at temperature zero with a fixed seed. This observation motivates assessing measurement variability through repeated queries; it does not establish independence across repetitions or a universal API-noise model.
}
may reduce observation noise, but it does not create additional independent respondents. 
Item calibration from tens or hundreds of distinct models can consequently be much less precise than calibration from large human testing samples.

At the same time, LLM evaluation can provide a type of information that is usually unavailable in conventional testing. 
For some model--item pairs, an evaluation system exposes a probability or likelihood associated with each answer option. 
Because LLM generation is based on a conditional distribution over the next token given the prompt and preceding tokens, the model assigns a probability to each candidate next token under fixed model parameters and context. 
When the evaluation protocol constrains the model to output an answer option immediately after the prompt, the probability assigned to the corresponding option token can be used as a model-implied probability for that option.
After normalization across the candidate options, the probability assigned to the correct option provides a fractional response in $[0,1]$. It records how much probability the model assigns to the answer, beyond the correctness of a single selected option.
Recent work has explored IRT-based and related methods for extracting more information from LLM benchmark responses \citep{polo2024,liao2025,fei2026,truong2026}.

However, probability responses are not universally available. 
Some LLMs do not expose token- or option-level probabilities through their application programming interfaces (APIs). 
Moreover, some reasoning-oriented LLMs perform intermediate reasoning before producing a final answer, making the correct-option probability unavailable as a direct next-token probability.

Restricting evaluation to probability responses therefore excludes models with binary-only records, whereas using only binary responses discards available probability information. 
Integrating the two response formats could expand the set of LLMs evaluated on a common scale without sacrificing the information contained in probability responses. 
This broader measurement coverage is valuable independently of whether integration improves item-calibration precision.
%This coverage is distinct from whether their observations improve item-calibration precision; meaningful common-scale comparisons still require appropriate measurement assumptions and aligned evaluation conditions.
%This issue is particularly relevant when combining results across evaluation systems or across generations of models: one source may provide correct-option probabilities, whereas another may provide only binary correctness. 
%Even when token probabilities are technically accessible, the probability assigned under one prompting or decoding protocol (the rule for selecting or sampling output tokens) need not coincide with the probability of producing a correct final answer under another. 
%Option normalization, answer extraction, prompting, and sampling determine whether probability and binary observations can reasonably be interpreted as measurements of the same underlying response tendency.
%
%Therefore, both response types can contribute to a common analysis only under appropriate alignment of their expected responses. Their different variances also make statistical efficiency a separate question: equal weights need not use the available measurements efficiently.
%

\paragraph{Purpose of this work}\mbox{}\

\begin{figure}[tbp]
\centering
\includegraphics[width=\linewidth]{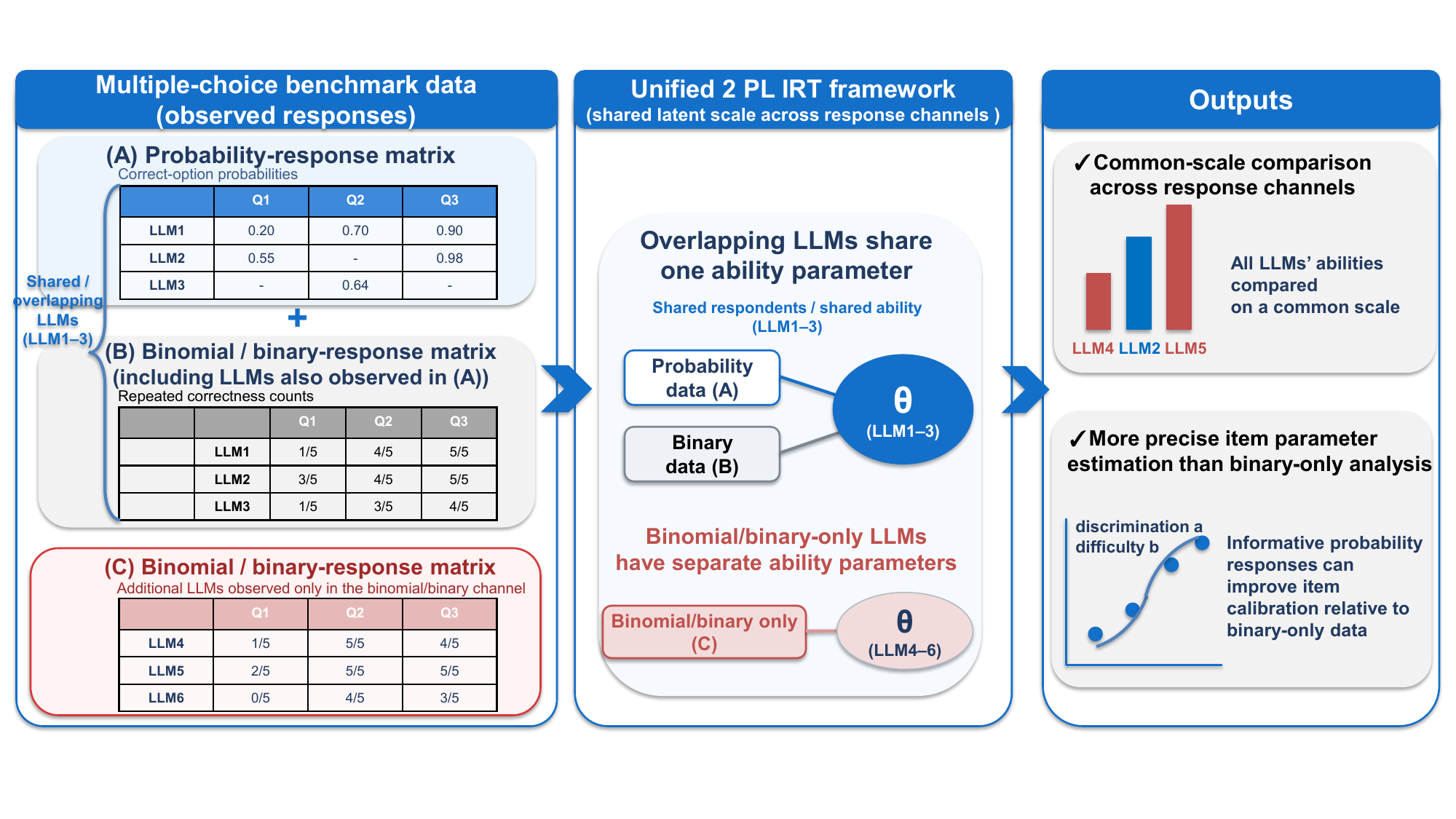}
\caption{Conceptual overview of the proposed IRT framework for integrating continuous probability responses and discrete binomial or binary responses, including shared latent abilities for overlapping LLMs.}
\label{fig: concept}
\end{figure}
In this work, we address this problem by formulating a common two-parameter logistic (2PL) IRT model, in which ability describes respondent performance, difficulty locates an item on that scale, and discrimination describes how sharply its expected response changes with ability. The model jointly accommodates probability and binary responses within a unified measurement framework,
as illustrated in Figure \ref{fig: concept}.
Probability observations enter through fractional-response quasi-maximum likelihood estimation (QMLE) \citep{papke1996}, while binary observations enter through the corresponding Bernoulli or binomial criterion. 
%Both response formats are linked to the same underlying conditional expectation, denoted by $\mu_{ij}$. 
%For a binary observation, $\mu_{ij}$ is the probability of a correct response, whereas for a probability observation, $\mu_{ij}$ is the conditional expectation of the observed correct-option probability. 
%When both formats are observed for the same LLM on the same item under compatible evaluation conditions, we assume that both observations reflect the same expected response and share the same latent ability parameter $\theta_i$.
%Because the Bernoulli likelihood and the fractional-response quasi-likelihood yield estimating equations centered on the same $\mu_{ij}$, the two response formats can be combined in a single estimating criterion.

Using numerical simulations, we further examine what can and cannot be gained by combining the two response channels in the proposed IRT framework.
Specifically, we study three distinct design questions: 1. how accurately item and ability parameters can be recovered, 2. how item uncertainty changes with the number of distinct LLM respondents, and 3. how ability uncertainty changes with the number of benchmark items. 
Here a response channel means one source of measurements in a specified format. We additionally distinguish adding new respondents from observing both channels for the same respondents.

Finally, we illustrate the framework using item-level MMLU~\citep{hendrycks2021} records from the Open LLM Leaderboard~\citep{openllm} and HELM~\citep{liang2023}. 
The analysis jointly incorporates probability-only, binary-only, and overlapping model entries within each subject. 
The analysis examines model-family response patterns and subject-level variation in item characteristics and dimensional structure. 
Simulation assesses statistical efficiency; the application examines measurement across heterogeneous leaderboard protocols.

\paragraph{Originality and scope of the proposed framework}\mbox{}\

Existing work uses IRT for model evaluation \citep{rodriguez2021}, adaptive testing \citep{zhuang2025}, item selection \citep{polo2024,kipnis2025}, continuous evaluation metrics \citep{liao2025}, and option-level probabilities \citep{fei2026,truong2026}. Human--AI comparisons and joint calibration also have established precedents \citep{lalor2016,gor2024caimira,liu2025respondents,ebara2026}. %Appendix~\ref{appendix: related work} reviews these related approaches.

Our contribution is the joint use of binary correctness and fractional correct-option probabilities in one 2PL estimating criterion, including partially overlapping respondents who share an ability parameter. Common-scale evaluation could also be obtained by fixing an item bank calibrated from probability responses and subsequently estimating abilities for binary-only models. The proposed joint framework instead accommodates binary-only, probability-only, and overlapping respondents in one estimation procedure without discarding probability information. Fractional-response QMLE already permits estimation without a full density model for fractional outcomes \citep{papke1996}. We use this established property for heterogeneous LLM evaluation records and examine the resulting precision with joint uncertainty calculations. The optimization and constrained covariance calculations use standard numerical methods. They are not separate methodological contributions.

We distinguish three issues throughout: compatibility of the estimating equations, statistical efficiency of integration, and adequacy of the measurement assumptions. Including models in this framework broadens the comparison set, but neither guarantees representativeness of the evolving, nonrandomly selected LLM population nor identifies its overall ability distribution. 
The simulations examine efficiency under a specified common response mean. The leaderboard application examines what happens when response formats and evaluation protocols differ in practice.

\section{Proposed Method: Integrated fractional-response IRT}
\subsection{Assumptions and algorithm for estimation}
Let $i=1,\ldots,N$ index distinct respondents (LLMs) and $j=1,\ldots,J$ index items. 
Let $a_j>0$ and $b_j$ denote item discrimination and difficulty, and let $\theta_i$ denote ability. 
The expected response at fixed ability and item parameters is
\begin{equation}\label{eq:mean}
 \mu_{ij}=\logitinv\{D a_j(\theta_i-b_j)\},\qquad D=1.702.
\end{equation}
For an observed binary channel, $K_{ij}$ is the number of correct answers in $R^B_{ij}$ trials. For an observed probability channel, $Q_{ijr}\in[0,1]$, $r=1,\ldots,R^P_{ij}$, is the reported probability of the correct answer. 
Following \citet{papke1996}, we assume that the two response types have the same expected response for given ability and item parameters
\footnote
{
Eq.~\eqref{eq:restriction} is a substantive alignment assumption.
For example, a token probability under one prompt need not equal the probability that a reasoning system produces the correct final answer under another protocol. 
Therefore, the two channels should be obtained under sufficiently comparable evaluation settings so that the responses can reasonably be interpreted as measurements of the same underlying response tendency.
This includes, for example, the prompt, answer-option normalization, answer extraction, and decoding or sampling procedure.
The simulations impose this alignment; they do not establish it for an actual API or benchmark. 
Likewise, the common item parameters assume measurement invariance across the respondent groups.
}
:
\begin{equation}\label{eq:restriction}
 \E(K_{ij}\mid\theta_i,a_j,b_j)=R^B_{ij}\mu_{ij},\qquad
 \E(Q_{ijr}\mid\theta_i,a_j,b_j)=\mu_{ij}.
\end{equation}
An unavailable channel has zero repeats; its count or sum contributes zero to the criterion.
The same $\theta_i$ appears in both restrictions when both channels are present, allowing both response types to contribute to the analysis of that LLM as a single respondent.

We rewrite the 2PL model as a logistic linear predictor, $\eta_{ij}=c_j\theta_i+\beta_j$, with $c_j=D a_j$ and $\beta_j=-c_jb_j$. This slope--intercept form simplifies the parameter updates. Because both response types yield the same score form, their contributions can be summed within each respondent--item pair:

\begin{equation}\label{eq:aggregate}
 Y_{ij}=K_{ij}+\sum_{r=1}^{R^P_{ij}}Q_{ijr},\qquad
 W_{ij}=R^B_{ij}+R^P_{ij}.
\end{equation}
We define the joint objective function as 
\begin{equation}\label{eq:objective}
 L(\theta,c,\beta)=\sum_{i,j}\left[W_{ij}\log(1+e^{\eta_{ij}})-Y_{ij}\eta_{ij}\right],\qquad c_j>0.
\end{equation}
We estimate the parameters by minimizing this objective function.
For the simulations, the initial parameter values $(\theta^{(0)},c^{(0)},\beta^{(0)})$ are obtained from the aggregated response data $(Y,W)$ as detailed in Appendix~\ref{subsec: initialization}. 
Every individual repeat has unit weight, prespecified independently of its measured value.

For binary counts, Eq. \eqref{eq:objective} is the binomial negative log likelihood up to a parameter-independent term. For fractional responses it is a working criterion, not a probability density. 
Its derivative with respect to a cell predictor is $W_{ij}\mu_{ij}-Y_{ij}$, whose expectation is zero under Eq. \eqref{eq:restriction}. 
This explains the compatibility of the scores without assigning Bernoulli variance to a fractional observation. 
We minimize the criterion with a standard block-coordinate scheme: update abilities with items fixed, update item slope--intercept pairs with abilities fixed, and normalize the scale. Joint IRT estimation uses such separate respondent and item updates \citep{chen2019}. Since $c_j\theta_i$ is a product of unknown parameters, the objective is not jointly convex, so convergence does not establish a global minimum.

Because the 2PL latent scale is invariant to affine transformations, the model requires location and scale constraints for identification. 
Specifically, the transformation
\begin{equation}\label{eq:transform}
 \theta_i^*=(\theta_i-m)/s,\quad c_j^*=s c_j,\quad
 \beta_j^*=\beta_j+c_jm,\qquad s>0,
\end{equation}
leaves every $\eta_{ij}$ unchanged. 
Therefore, we impose two normalization constraints during estimation, using either item-based constraints or reference-respondent constraints to define the scale.
These identification constraints fix the origin and unit of the latent scale without changing fitted response probabilities. Section~\ref{sec:design} specifies the constraints for each simulation target.

Algorithm~\ref{alg:fit} summarizes the process of estimation. 
Appendix~\ref{subsec: initialization} describes the numerical updates and convergence checks for the simulations. Appendix~\ref{sec:empirical-computation} describes the empirical estimation procedure. 
%The simulations use this unpenalized estimator. 
%The empirical version adds a weak ridge penalty and allows zero slopes to stabilize estimation for weakly discriminating items and nearly separated response patterns.

\begin{algorithm}[t]
\caption{Fractional-response IRT: conceptual alternating minimization}\label{alg:fit}
\begin{algorithmic}[1]
\Require Aggregated response data $(Y,W)$ from Eq.~\eqref{eq:aggregate}; identification constraints
\State Initialize $(\theta^{(0)},c^{(0)},\beta^{(0)})$, $c^{(0)}>0$; normalize to the constraints by Eq. \eqref{eq:transform}.
\For{$t=0,1,\ldots$ until convergence}
  \State $\displaystyle \theta^{(t+1)}\gets\operatorname*{arg\,min}_{\theta} L(\theta,c^{(t)},\beta^{(t)})$
  \State $\displaystyle (c^{(t+1)},\beta^{(t+1)})\gets\operatorname*{arg\,min}_{c>0,\,\beta} L(\theta^{(t+1)},c,\beta)$
  \State Normalize $(\theta^{(t+1)},c^{(t+1)},\beta^{(t+1)})$ to the constraints by Eq. \eqref{eq:transform}, preserving $\eta_{ij}$.
\EndFor
\State \Return $\hat\theta_i=\theta_i^{(t+1)}$, $\hat a_j=c_j^{(t+1)}/D$, $\hat b_j=-\beta_j^{(t+1)}/c_j^{(t+1)}$.
\end{algorithmic}
\end{algorithm}

\subsection{Analytical standard errors}\label{sec:sandwich}
We calculate standard errors (SEs) analytically using a constrained cluster-score sandwich covariance estimator \citep{white1982}, referred to below as ``analytical SEs''. 
This is a sampling-variance approximation, not an exact finite-sample result. 
The calculation uses the full identified parameter vector $\psi=(\theta,\alpha,\beta)$, where $\alpha_j=\log c_j$. 
Let $H=\nabla^2_\psi L(\hat\psi)$ be the observed Hessian of the joint objective function in Eq.~\eqref{eq:objective}. 
Let $h(\psi)=0$ denote the identification constraints, and let
$C=\partial h(\psi)/\partial\psi^\top|_{\hat\psi}$ be their Jacobian.
Write the joint objective as $L(\psi)=\sum_{k=1}^K L_k(\psi)$ over clusters,
and define the cluster score
\[
s_k=-\nabla_\psi L_k(\hat\psi).
\]
Let $S=[s_1,\ldots,s_K]$ collect these cluster scores.
Let $U=[u_1,\ldots,u_K]$ collect the constrained influence vectors
associated with the cluster scores, and let $V$ contain the corresponding
auxiliary constraint multipliers. The influence vectors and multipliers are obtained by solving the standard Karush--Kuhn--Tucker (KKT) linear system for constrained influence vectors:
\[
\begin{pmatrix}
H & C^\top\\
C & 0
\end{pmatrix}
\begin{pmatrix}
U\\
V
\end{pmatrix}
=
\begin{pmatrix}
S\\
0
\end{pmatrix}.
\]
Solving this system maps each cluster score $s_k$ to its corresponding first-order influence $u_k$ on the constrained estimator. 
Therefore, the analytical covariance estimate is
\begin{equation}\label{eq:border}
\qquad
 \widehat{\operatorname{Cov}}(\hat\psi)=\frac{K}{K-1}UU^\mathsf{T}.
\end{equation}
For item SEs, a cluster is a whole respondent row ($K=N$), and different respondents are assumed independent. For ability SEs, a cluster is a whole item column ($K=J$), and different items are assumed independent. Responses within a cluster may be dependent, including paired probability and binary records. Related LLMs and dependent benchmark items can violate these assumptions; clustering alone does not resolve dependence between clusters.

Then, the analytical SE for each component $\hat\psi_r$ is obtained as
\begin{equation}
\widehat{\mathrm{SE}}(\hat\psi_r)
=
\sqrt{
\left[
\widehat{\operatorname{Cov}}(\hat\psi)
\right]_{rr}
}.
\label{eq:SE_cov}
\end{equation}

Because the covariance is calculated in the estimation parameterization
$(\theta,\alpha,\beta)$, analytical SEs for the reported parameters
$(\theta,a,b)$ are obtained by the delta method. 
Since
$a_j=e^{\alpha_j}/D$ and $b_j=-\beta_j/e^{\alpha_j}$,
\[
\widehat{\mathrm{SE}}(\hat a_j)
=
\hat a_j\,\widehat{\mathrm{SE}}(\hat\alpha_j),
\]
and
\[
\widehat{\mathrm{SE}}(\hat b_j)
=
\left[
\hat b_j^2\widehat{\mathrm{Var}}(\hat\alpha_j)
+
\frac{\widehat{\mathrm{Var}}(\hat\beta_j)}{\hat c_j^2}
+
\frac{2\hat b_j}{\hat c_j}
\widehat{\mathrm{Cov}}(\hat\alpha_j,\hat\beta_j)
\right]^{1/2}.
\]
The SEs of $\theta_i$ are obtained directly from the corresponding
diagonal elements of $\widehat{\operatorname{Cov}}(\hat\psi)$.

\section{Simulation design}\label{sec:design}
We perform simulations to evaluate the estimator, assuming that the probability and binary responses share the specified 2PL mean. 
We examine whether the parameters are recovered, whether the uncertainty estimates describe sampling variability, and whether adding respondents with a different response type improves precision. 
%They do not test whether the common-mean assumption holds for actual leaderboard protocols.
Appendix~\ref{sec:detailed design} gives the response-generation and numerical estimation details.

\subsection{Data generation and measurement conditions}

We first generate the true item and ability parameters independently, with
$a_j$ and $b_j$ drawn from uniform distributions and $\theta_i$ from a
standard normal distribution. Each available response channel has five repeats.
Binary responses are generated as
$K_{ij}\sim\mathrm{Binomial}(5,\mu_{ij})$.
Probability responses $Q_{ijr}\in[0,1]$ are generated from a beta distribution
with mean $\mu_{ij}$ and variance lower than that of a binary trial with the same mean. This variance is a simulation design choice; it is not an empirically calibrated model of API noise.
Given the generated parameters, we draw responses
independently across repeats, response channels, and model--item cells.

We compare four measurement conditions:
(A) binary responses only;
(B) probability responses only;
(C) one channel per respondent, with approximately half of the respondents observed through probability responses and the remainder through binary responses
(probability/binary split); and
(D) both channels per respondent, with the two response channels sharing the same ability parameter for each respondent (both channels).
The numbers of respondents and items vary across experiments as described below.

\subsection{Types of experiments}
\begin{table}[!t]
\centering\small
\caption{Types of simulation experiments. Each condition has 100 Monte Carlo replicates. SE experiments attempt 200 bootstrap draws per eligible dataset. 
%Both size grids span 25--1,000.
}\label{tab:design of experiments}
\begin{tabular}{p{.23\linewidth}p{.35\linewidth}p{.32\linewidth}}
\toprule
Experiment type & Design & Primary target \\
\midrule
1. Parameter recovery & $N=100$, $J=100$; both banks regenerated & RMSE of $a,b,\theta$ \\
2. Respondent-count dependence of item uncertainty & $N=25$--$1000$; fixed $J=100$ & RMSSE of $a, b$ \\
3. Item-count dependence of ability uncertainty & $J=25$--$1000$; fixed $N=100$ & RMSSE of $\theta$ \\
4. Adding probability respondents & 100 binary respondents plus 0--1000 probability respondents; $J=100$ & RMSE and RMSSE of $a,b$ and the original binary respondents' $\theta$ \\
5. Adding binary respondents & 100 probability respondents plus 0--1000 binary respondents; $J=100$ & RMSE and RMSSE of $a,b$ and the original probability respondents' $\theta$ \\
\bottomrule
\end{tabular}
\end{table}
Table~\ref{tab:design of experiments} summarizes the designs.
The experiments distinguish parameter recovery, uncertainty scaling, and the effect of adding respondents with a different response format.
Each experimental condition is evaluated using 100 Monte Carlo replicates\footnote{
For item inference, we impose $J^{-1}\sum_j\log c_j=0$ and $J^{-1}\sum_j\beta_j=0$.
For ability inference, the reference abilities are normalized to have zero mean and unit second moment. 
These constraints define the latent scale and do not treat the reference parameters as known.
}
.

\paragraph{Parameter recovery}\mbox{}\

We first examine whether the proposed estimation procedure accurately recovers the true parameter values. Estimation accuracy is compared across conditions (A)--(D) using the root mean squared error (RMSE),
\begin{equation}
 \RMSE=
 \left\{
 \frac{1}{Md}
 \sum_{m=1}^M
 \sum_{h=1}^d
 (\hat\xi_{mh}-\xi_{mh})^2
 \right\}^{1/2}
 \label{eq:RMSE}
\end{equation}
where $M$ is the number of simulation replicates and $d$ is the number of
target parameters.

\paragraph{Respondent-count dependence of item uncertainty}\mbox{}\

We vary the number of respondents $N$ and examine how the uncertainty of
the item-parameter estimates changes with sample size. 
The overall magnitude of the estimated standard errors is summarized by the root mean squared standard error (RMSSE),
\begin{equation}
 \RMSSE=
 \left\{
 \frac{1}{Md}
 \sum_{m=1}^M
 \sum_{h=1}^d
 \widehat{\mathrm{SE}}_{mh}^{\,2}
 \right\}^{1/2},
 \label{eq:RMSSE}
\end{equation}
where $\widehat{\mathrm{SE}}_{mh}$ is obtained either analytically from Eqs.~(\ref{eq:border}) and (\ref{eq:SE_cov}) or from a pairs bootstrap based on 200 resamples\footnote{
Analysis of each bootstrap resample uses the same identification and normalization constraints as the corresponding analysis on Monte Carlo replicates.
}.

To assess whether these estimated standard errors reflect the actual sampling variability, we compare RMSSE with the empirical standard deviation (SD) across Monte Carlo replicates. 

\paragraph{Item-count dependence of ability uncertainty}\mbox{}\

To check the expected effect of test length on ability precision, we vary the number of items $J$ and examine how the uncertainty of the ability estimates changes with test length. 
As in the respondent-count experiment, analytical RMSSE and bootstrap RMSSE are compared with the empirical SD across Monte Carlo replicates.

\paragraph{Adding respondents to a fixed cohort}\mbox{}\

We also examine whether increasing the number of respondents necessarily improves estimation when response precision differs between groups. With $J=100$, we add probability-only respondents to 100 binary-only respondents, and separately add binary-only respondents to 100 probability-only respondents. The original respondents and all their observations stay fixed within each replicate. Item errors concern a fixed item bank; ability errors concern only the original 100 respondents.

At each design point, we also evaluate binary-only and probability-only
controls using the same respondent identities and item bank.

\section{Results of simulation}
\subsection{Parameter recovery}
\begin{figure}[!htbp]
\centering\includegraphics[width=\linewidth]{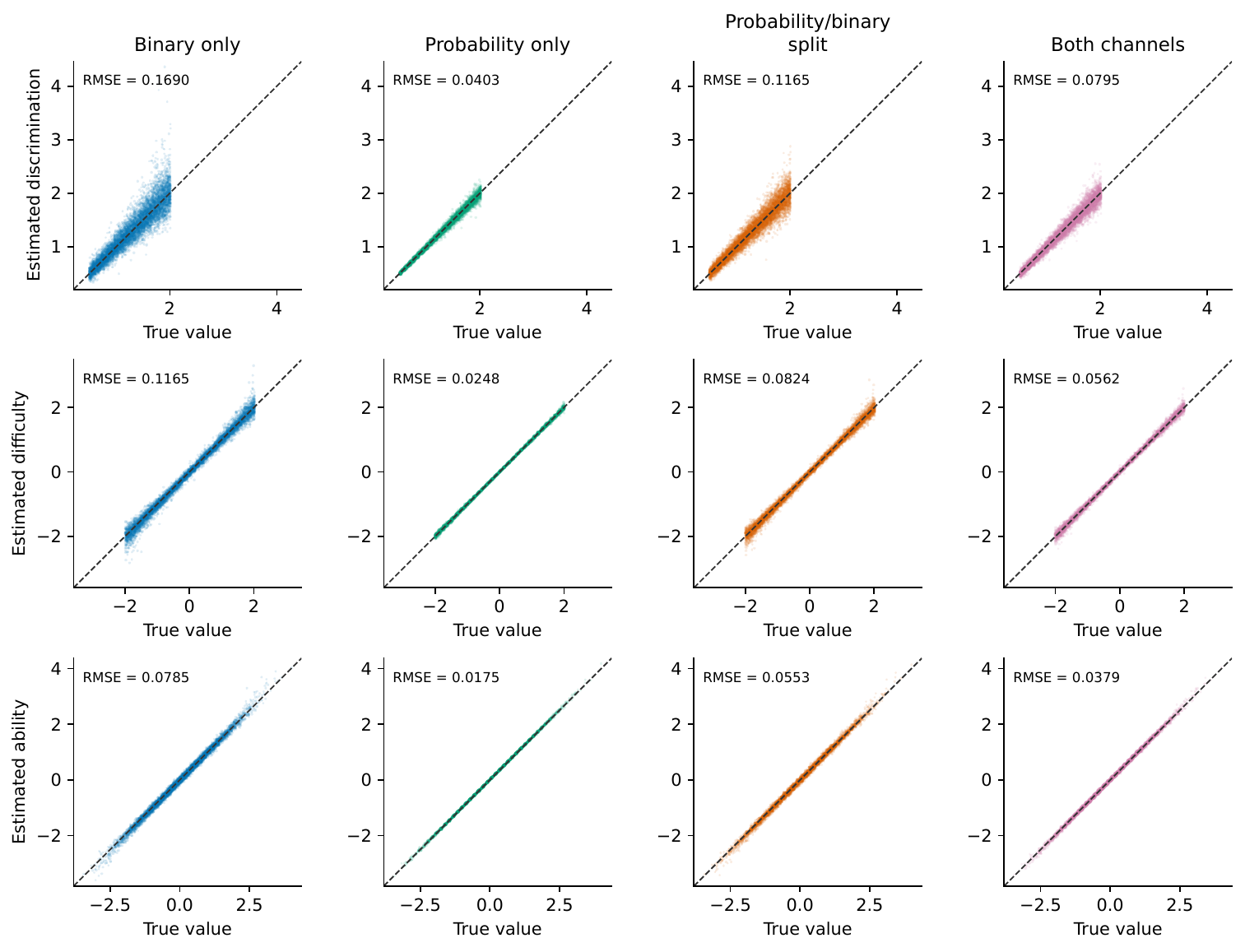}
\caption{
True versus estimated parameters under the primary generator ($N=J=100$). Each panel includes 10,000 parameter coordinates from all 100 Monte Carlo replicates. 
The dashed line, $y=x$, represents exact recovery of the generating values.  
}
\label{fig:recovery-scatter}
\end{figure}
Figure~\ref{fig:recovery-scatter} shows recovery of the generating parameters under all four measurement conditions. Probability-only responses give the most precise estimates. %This is a validation check for the estimator under the specified mean model.

Parameter recovery is more precise with the probability/binary split than with binary-only measurement, but less precise than with probability-only measurement. Observing both channels for every respondent further improves recovery relative to the split condition, but still yields larger RMSE than probability-only measurement, despite the additional observations.

These results demonstrate the feasibility of integrating binary and probability responses within a common IRT framework. 
Under the common conditional-mean assumption, probability responses bounded in $[0,1]$ have conditional variance no greater than Bernoulli responses with the same mean. 
This lower response variability reduces fluctuations in the estimating scores and, under otherwise identical measurement conditions, contributes to more accurate parameter recovery. 
However, integrating additional binary observations does not necessarily reduce RMSE under equal observation weights, as further examined in Sections \ref{subsec: adding p} and \ref{subsec: adding b}.

\subsection{Respondent-count dependence of uncertainty of item parameters}
\begin{figure}[!t]
\centering\includegraphics[width=\linewidth]{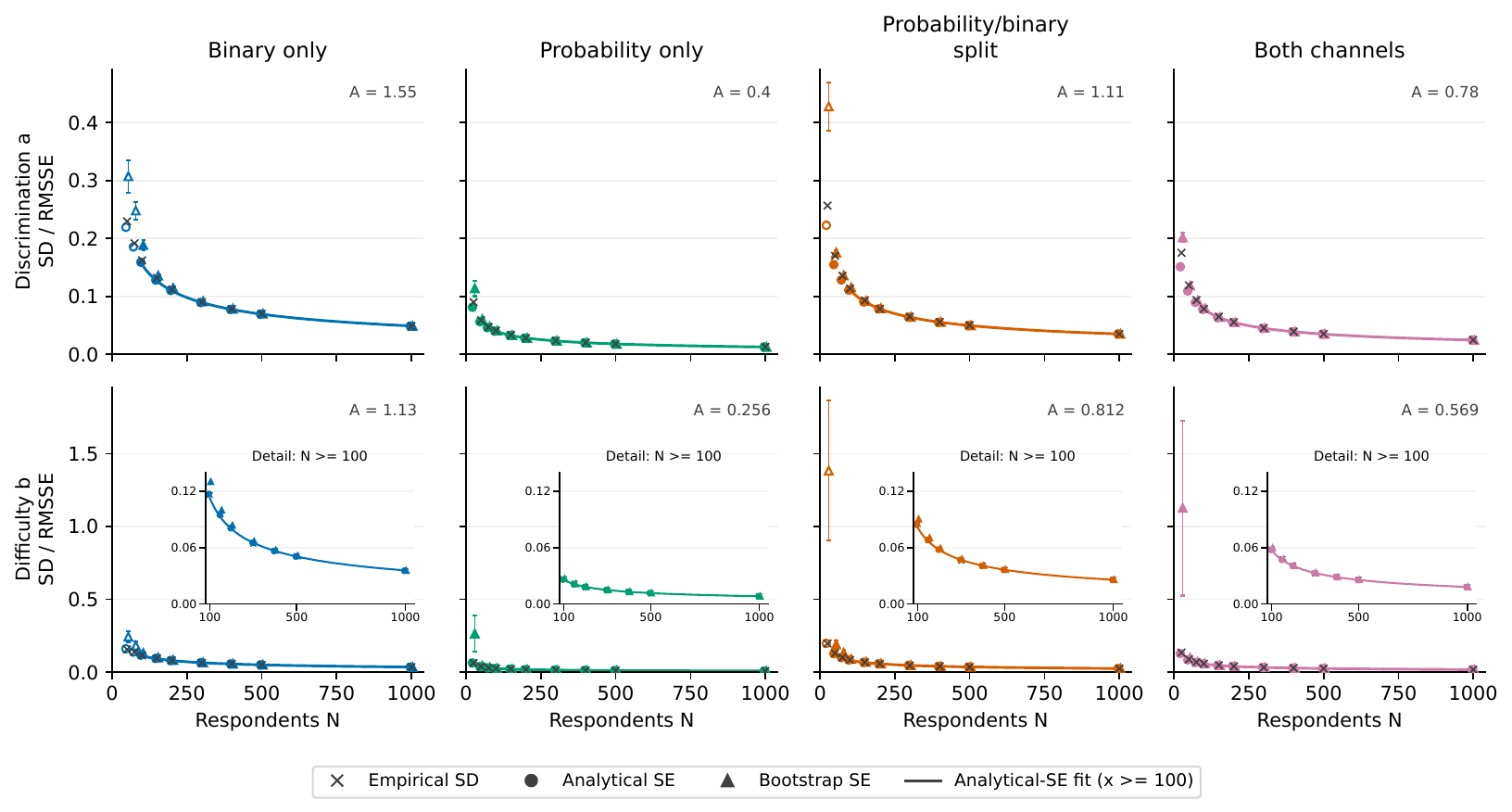}
\caption{Item uncertainty (RMSSE) as respondent count varies ($J=100$). 
Unconnected markers show empirical SD, analytical RMSSE, and bootstrap RMSSE for the same retained datasets.
Error bars summarize Monte Carlo uncertainty as $\pm1.96$ Monte Carlo standard errors. 
Analytical RMSSEs are fitted with curves of the form $A/\sqrt{N}$ for $N\geq100$. 
Difficulty insets enlarge this domain. 
}
\label{fig:sample}
\par\medskip
\centering\includegraphics[width=\linewidth]{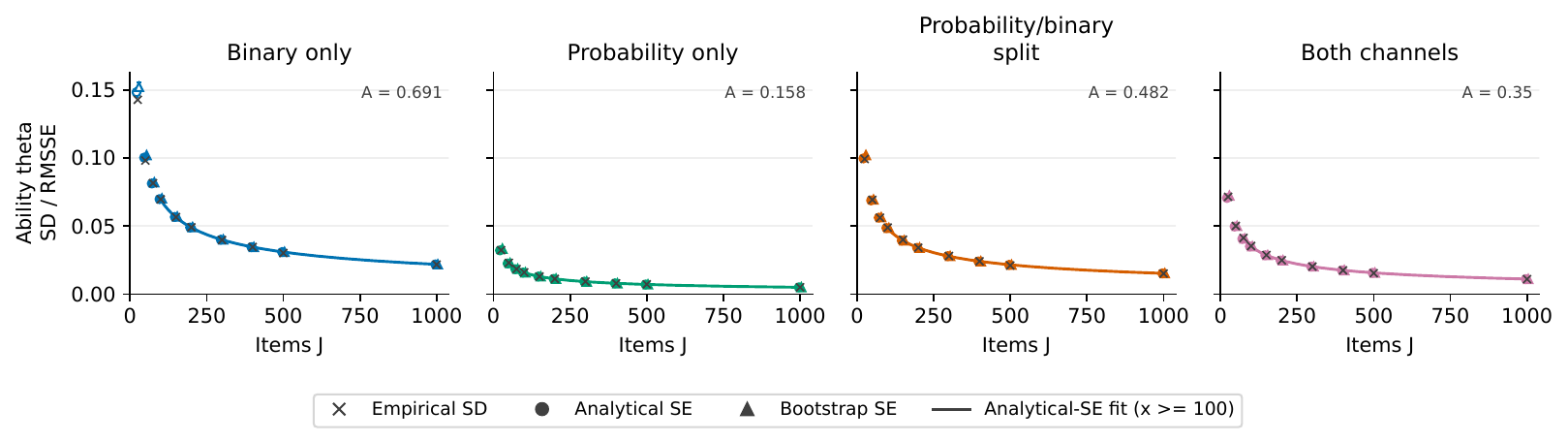}
\caption{
Ability uncertainty (RMSSE) as item count varies ($N=100$). 
Unconnected markers show empirical SD, analytical RMSSE, and bootstrap RMSSE for the same retained datasets.
Error bars summarize Monte Carlo uncertainty as $\pm1.96$ Monte Carlo standard errors.
Analytical RMSSEs are fitted with curves of the form $A/\sqrt{J}$ for $J\geq100$.
}
\label{fig:items}
\end{figure}
Figure~\ref{fig:sample} shows how the uncertainty of the item parameters changes with the number of respondents $N$. 
Across all four measurement conditions, analytical RMSSE and bootstrap RMSSE closely agree with the empirical SD across Monte Carlo replicates, indicating that the estimated standard errors adequately capture the observed sampling variability.

To quantify the dependence on sample size, we fit the analytical RMSSE for $N \geq 100$ using $A/\sqrt{N}$, where $A$ represents the magnitude of the uncertainty of estimation at a given number of respondents. 

The fitted curves closely approximate the observed decrease in RMSSE, showing a common $N^{-1/2}$ scaling across response formats.
However, $A$ differs substantially between the measurement conditions. 
The probability-only condition yields the smallest uncertainty at a given number of respondents, while the binary-only condition yields the largest. 
The two integrated conditions lie between these extremes, with both-channel measurement providing lower uncertainty than the split condition.

These differences are consistent with the lower conditional variability of probability responses under the common conditional-mean assumption. 
As discussed in Section \ref{sec:sandwich}, response variability enters the score-variance component of the sandwich covariance. 
Consequently, item calibration can be more precise with probability responses than with binary responses.
In other words, a comparable level of item-parameter uncertainty can be achieved with fewer probability respondents than binary respondents. 

\subsection{Item-count dependence of ability uncertainty}

Figure~\ref{fig:items} shows that uncertainty in ability estimates decreases as the number of items $J$ increases. 
Both analytical and bootstrap RMSSE agree closely with the empirical SD across Monte Carlo replicates, supporting the reliability of the proposed analytical SEs for ability estimation. Furthermore, the decrease in RMSSE is well approximated by $A/\sqrt{J}$ over the examined item-count range, consistent with the expected $J^{-1/2}$ scaling of the estimation uncertainty.

Probability-only measurement consistently yields the smallest RMSSE, while binary-only measurement yields the largest. 
This indicates that, under the specified low-variance probability-response mechanism, probability responses can achieve a comparable level of ability precision with fewer items than binary responses. 
The integrated measurement conditions lie between these two extremes, consistent with the parameter-recovery experiment.

\subsection{Adding probability respondents to a fixed binary cohort}\label{subsec: adding p}
\begin{figure}[!t]
\centering\includegraphics[width=\linewidth]{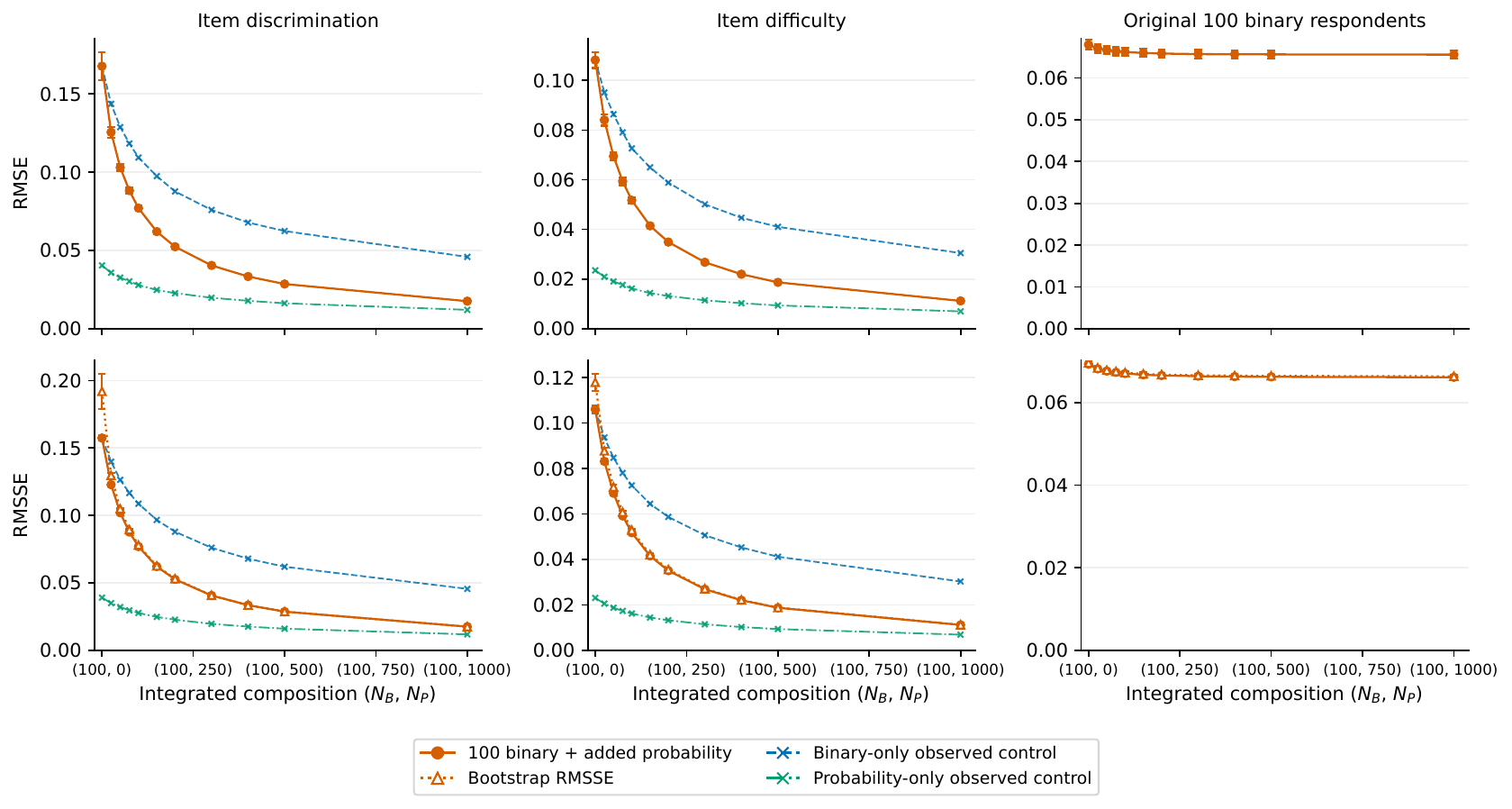}
\caption{Adding probability respondents to 100 fixed binary respondents ($J=100$). Each design point is identified by $(N_B,N_P)$. 
Upper panels show RMSE; lower panels show analytical and bootstrap RMSSE. 
Bars summarize Monte Carlo uncertainty as $\pm1.96$ Monte Carlo standard errors. 
Item controls use the same identities and item bank with all responses measured in one format.
Ability panels concern only the original binary respondents.}
\label{fig:add-probability}
\par\medskip
\centering\includegraphics[width=\linewidth]{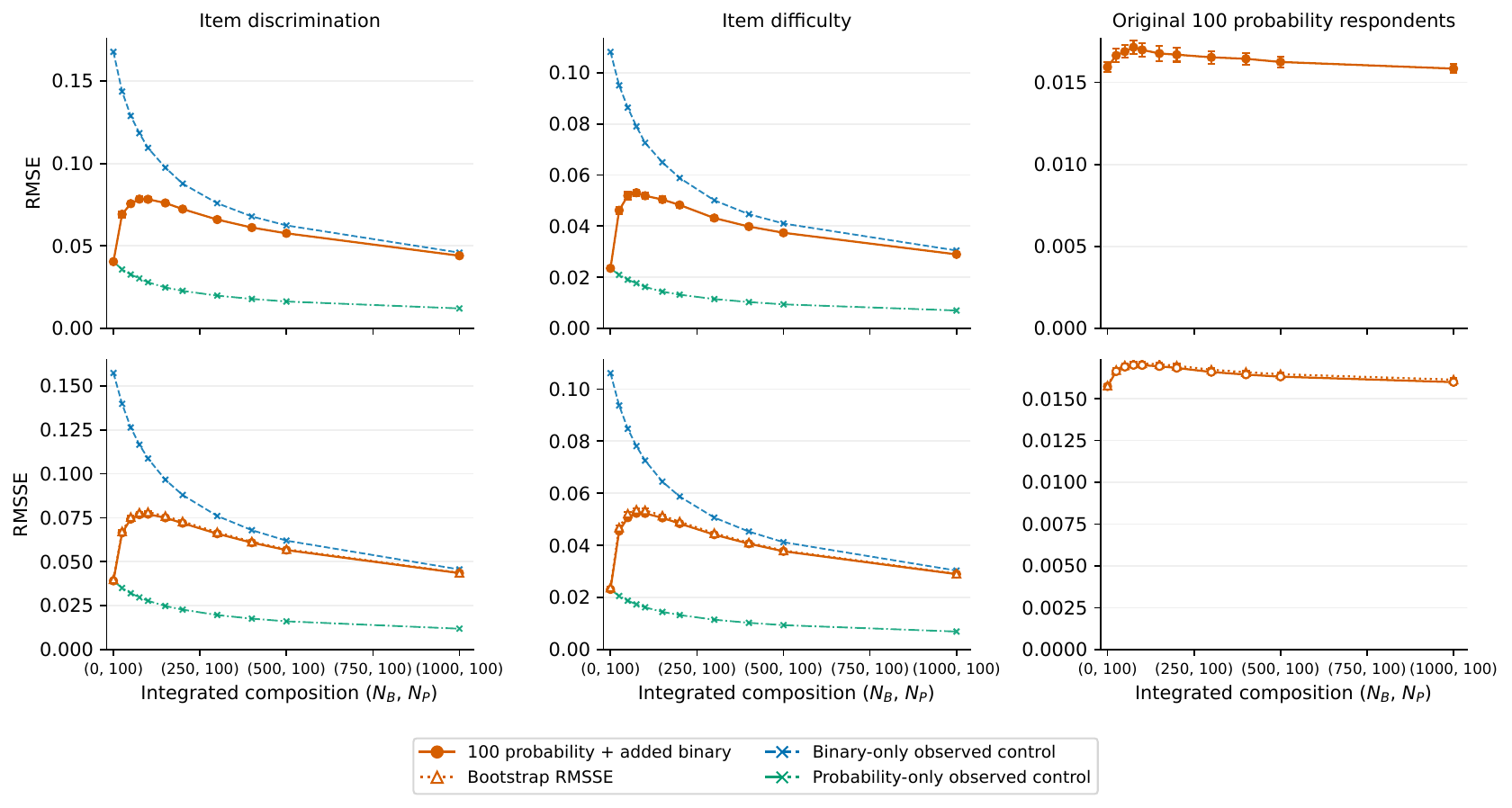}
\caption{Adding binary respondents to 100 fixed probability respondents ($J=100$). Design points are labeled by $(N_B,N_P)$. Markers, observed single-format controls, and Monte Carlo bars follow Figure~\ref{fig:add-probability}. Ability panels concern only the original probability respondents. 
%The response composition changes along the horizontal axis; the mixed results are not fitted by a function of total respondent count.
}
\label{fig:add}
\end{figure}
Figure~\ref{fig:add-probability} shows that adding probability respondents steadily improves item calibration. 
The item curves move from the binary-only baseline toward the probability-only reference as probability records become more prevalent. 
They remain distinct from probability-only calibration because the original binary records still contribute to the estimating equation.

Adding probability respondents improves item-calibration precision more than adding the same number of binary respondents. 
Notably, adding 50 probability respondents to the original 100 binary respondents yields slightly smaller item-parameter RMSSE than doubling the binary sample to 200 respondents.

In contrast, the ability estimates of the original binary respondents show only modest improvements in precision, as their own responses and the number of answered items remain unchanged. 
The benefit of adding probability respondents is therefore more pronounced for item calibration than for the existing respondents' ability estimates.

\subsection{Adding binary respondents to a fixed probability cohort}\label{subsec: adding b}

Figure~\ref{fig:add} shows that adding binary respondents to an existing probability sample produces a different pattern. 
The RMSE and RMSSE of the item parameters initially increase, indicating that additional binary observations can reduce item-calibration precision under the equal-weight estimation procedure, even when both response types share the same conditional mean.

However, as more binary respondents are added, the RMSE and RMSSE of the item parameters begin to decrease, approaching the binary-only reference as binary observations become dominant.

As in the preceding experiment on adding probability respondents, the uncertainty of the original respondents' ability estimates changes relatively little compared with the uncertainty in item parameters.

Together, the two experiments demonstrate an asymmetric effect of adding heterogeneous response data: adding probability respondents to a binary sample substantially improves item-calibration precision, whereas adding binary respondents to a probability sample can initially worsen it. 
Thus, the effect of integration on the precision of the estimation depends on the response formats and composition of the existing and additional samples.

\section{Empirical application: integrating MMLU leaderboards}\label{sec:empirical}
The application to real-world MMLU response datasets addresses four questions: 
how IRT ability differs from average correct-option probability; 
whether the two sources yield comparable abilities and how integration changes uncertainty; 
how item characteristics vary across subjects; 
and whether a one-dimensional model is adequate within each subject. 
The application illustrates joint estimation for probability-only, binary-only, and overlapping entries; it does not establish measurement equivalence between the source protocols. 
Throughout this section, \emph{Open} denotes Open LLM Leaderboard probability records and \emph{HELM} denotes binary records. 
These names identify the sources; probability and binary identify their response formats. 
\emph{Integrated} denotes use of both sources with a shared ability for overlapping entries.

\subsection{Data sources and response formats}
The empirical application combines item-level MMLU evaluations from Hugging Face's Open LLM Leaderboard and Stanford's Holistic Evaluation of Language Models (HELM). 
MMLU comprises 57 subjects and 14,042 test questions \citep{hendrycks2021}. 
The archived Open LLM Leaderboard evaluated MMLU with five-shot prompts and provides detailed model-level evaluation records \citep{openllm}; 
HELM provides a framework and public evaluation records of open and restricted-access models \citep{liang2023}. 
The HELM runs use a different prompting format, temperature-zero decoding, and a one-token generation limit. 
Their recorded exact-match outcomes are therefore used here as binary correctness observations.

The Open LLM Leaderboard records instead provide likelihoods for the candidate answer options. 
After normalization across candidates, the probability assigned to the correct option is used as the probability response.

Although the two sources do not use identical evaluation protocols,
\footnote{
The two protocols should not be regarded as equivalent merely because both use five-shot prompting. 
Appendix~\ref{sec:protocol} describes the protocol differences and their implications for comparing the responses.
}
we jointly analyze the two response formats using the proposed framework as an empirical application, while acknowledging that differences in evaluation protocols may violate the common-mean assumption.

Moreover, different MMLU subjects can measure different capabilities. 
We fit a separate IRT model in each subject to avoid placing all questions in different MMLU subjects on a single ability axis.

Table~\ref{tab:empirical-cohorts} describes the source snapshot before applying the empirical inclusion and exclusion criteria in Appendix~\ref{sec:empirical-computation}. 
The snapshot includes 6,465 Open LLM Leaderboard-only model entries, 57 HELM-only entries, and 22 entries present in both sources: 6,544 distinct entries in total\footnote{
These counts were recomputed from rows containing at least one observed response across the 57 archived matrices per channel. 
More precisely, subject-specific respondent totals range from 6,537 to 6,544 because channel availability varies. 
These are snapshot counts, not current leaderboard totals or counts of independently developed model families. Appendix~\ref{sec:protocol} lists the shared model identifiers. 
}.  
In the 22 overlapping cases, we assume that the two channels share one ability parameter within each subject.

The substantially larger Open LLM Leaderboard sample provides most of the observations for item calibration. 
Nevertheless, integrating HELM allows additional models, including \texttt{gpt-4-1106-preview}, \texttt{claude-3-sonnet-20240229}, and \texttt{gemini-1.0-pro-001}, to be evaluated within the same IRT framework. 
This illustrates the value of integration in extending model coverage, even when gains in item-calibration precision are limited.

\begin{table}[t]
\centering\small
\caption{Model entries in the integrated MMLU analysis. 
Each available channel contributes one stored observation per observed model--item pair. 
Each of the 22 overlapping models is represented by one shared ability per subject.
}\label{tab:empirical-cohorts}
\begin{tabular}{lrl}\toprule
Data source leaderboard & Number of models & Response format \\
\midrule Open LLM Leaderboard only & 6,465 & probability \\
HELM only & 57 & binary \\
Both leaderboards (shared ability) & 22 & probability and binary \\
\midrule Total distinct model entries & 6,544 & \\
\bottomrule

\end{tabular}
\end{table}

\subsection{Empirical estimation}
\begin{samepage}
We apply the empirical version of Algorithm~\ref{alg:fit}, with the weak ridge penalty and zero-slope treatment specified in Appendix~\ref{sec:empirical-computation}, to stabilize estimation when response patterns provide limited information about item parameters. 
For this application, $q_{ij}$ is the continuous correct-option probability used in the Open LLM Leaderboard analysis, and $z_{ij}\in\{0,1\}$ is the binary correctness indicator from HELM. 
There is one stored observation per available channel. 
With $\mathcal O_P$ and $\mathcal O_B$ denoting the observed pairs in each channel, the empirical criterion is
\begin{align}\label{eq:empirical}
 L_{\mathrm{emp}}={}&-\sum_{(i,j)\in\mathcal O_P}
  \{q_{ij}\log\mu_{ij}+(1-q_{ij})\log(1-\mu_{ij})\}\nonumber\\
 &-\sum_{(i,j)\in\mathcal O_B}
  \{z_{ij}\log\mu_{ij}+(1-z_{ij})\log(1-\mu_{ij})\}.
\end{align}
\end{samepage}
This objective function corresponds to Eq.~\eqref{eq:objective} for a single observation.

Within each subject, the ability scale has mean zero and standard deviation one among the eligible Open LLM Leaderboard reference respondents.
The resulting ability scales are subject-specific and are not directly comparable across subjects. 
Analytical SEs are computed by extending the method in Section~\ref{sec:sandwich} to the penalized objective, with the SEs transformed to the standardized ability scale. 
The calculation treats the fitted set of zero-slope items as fixed during differentiation, as detailed in Appendix~\ref{sec:empirical-computation}.

\subsection{Ability estimates and model-family response patterns}
\begin{figure}[t]
\centering\includegraphics[width=\linewidth]{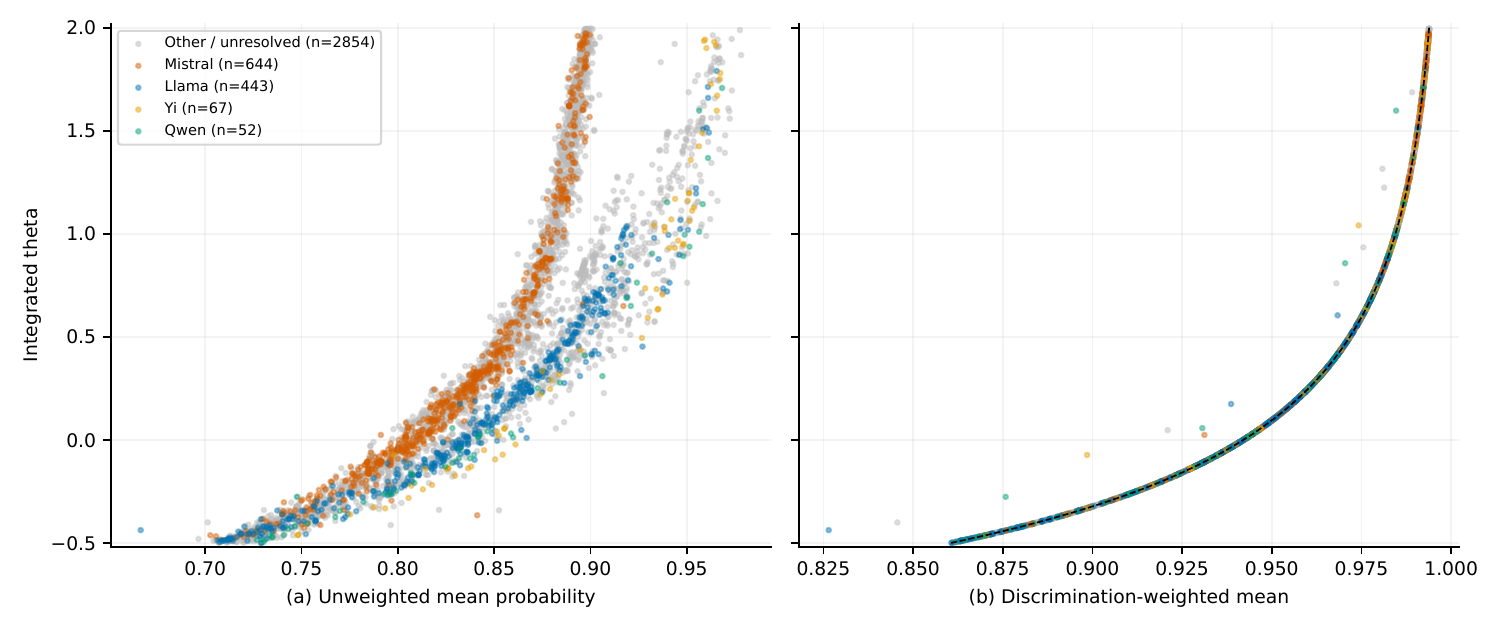}
\caption{
Estimated abilities in high school government and politics, the subject with the highest mean discrimination in Table~\ref{tab:empirical-domains}, against mean correct-option probabilities: (a) unweighted and (b) weighted by item discrimination $a$. 
The displayed range, $-0.5< \theta \leq 2$, highlights the curved bands of points, called branches here.
Colors indicate provisional model-family labels derived from metadata and model names. 
The dashed curve in (b) is the model-implied weighted mean, not a fitted spline.}
\label{fig:mistral-branch}
\end{figure}

Figure~\ref{fig:mistral-branch} relates the estimated ability to mean correct-option probability in high school government and politics, the subject with the highest mean discrimination in Table~\ref{tab:empirical-domains}. The six subjects in that table---those with the three highest and three lowest mean discrimination values---also appear in Table~\ref{tab:dim-examples}, with two of the six illustrated in Figure~\ref{fig:empirical-dim}. Only probability-response records are shown here.

As expected, the two variables are positively associated. 
However, the relationship is not represented by a single branch: several curved bands of points, or branches, appear among models with similar average probabilities.
To explore this behavior, the observations are colored by model family.
In particular, the upper branch is populated predominantly by models derived from Mistral, suggesting that the branching pattern is associated with systematic differences among model families.

Figure~\ref{fig:mistral-branch}(b) replaces the unweighted mean probability with the mean weighted by the estimated item discrimination $a_j$. 
The two branches then nearly coincide along a single curve. This pattern can be explained by the relationship between the discrimination-weighted mean response and the estimated ability under the 2PL model.
Differentiating the objective function in Eq.~\eqref{eq:empirical} with respect to $\theta_i$ gives the ability score equation. 
For a probability-only respondent observed on a common item set, the unpenalized first-order condition is
\[
\sum_j a_j(q_{ij}-\mu_{ij})=0.
\]
In the empirical analysis, the weak ridge penalty modifies this first-order condition, so the equality holds approximately.
Rearranging gives
\[
\frac{\sum_j a_jq_{ij}}{\sum_j a_j}
\approx
\frac{\sum_j a_j\mu_j(\hat\theta_i)}{\sum_j a_j}.
\]
The right-hand side is strictly increasing in $\hat\theta_i$ when item parameters are fixed and at least one discrimination parameter is positive. 
Thus, the observed discrimination-weighted mean is closely constrained by the estimated ability.

In contrast, no equivalent constraint holds for the unweighted mean response.
Although its expected value also increases with ability under the 2PL model, the unweighted residuals need not sum to zero, allowing models with similar estimated abilities to have different unweighted mean probabilities. 
This explains why the branches can remain visible in Figure~\ref{fig:mistral-branch}(a) but largely disappear after discrimination weighting in panel (b). 
The near-coincidence therefore reflects the score equation underlying ability estimation, rather than independent evidence that model-family differences have disappeared.

\subsection{Overlapping models: channel drift and uncertainty}

The 22 overlapping models allow a comparison of Open-only, HELM-only, and integrated abilities while holding the integrated item parameters fixed. 
We call the resulting uncertainty estimates \emph{fixed-item SEs}: they describe ability uncertainty for a given calibrated bank and do not include the uncertainty of the item estimates.

The point estimates exhibit a systematic channel shift in ability.
HELM binary responses tend to produce higher ability estimates than the Open probability responses, and the combined estimates consequently tend to shift upward when the HELM channel is added.
This difference is closely related to the different response definitions used by the two leaderboards. 
Open records provide normalized candidate-option likelihoods, whereas HELM uses generated answers and records binary correctness at zero temperature.

\begin{figure}[tbp]
\centering
\includegraphics[width=\linewidth]{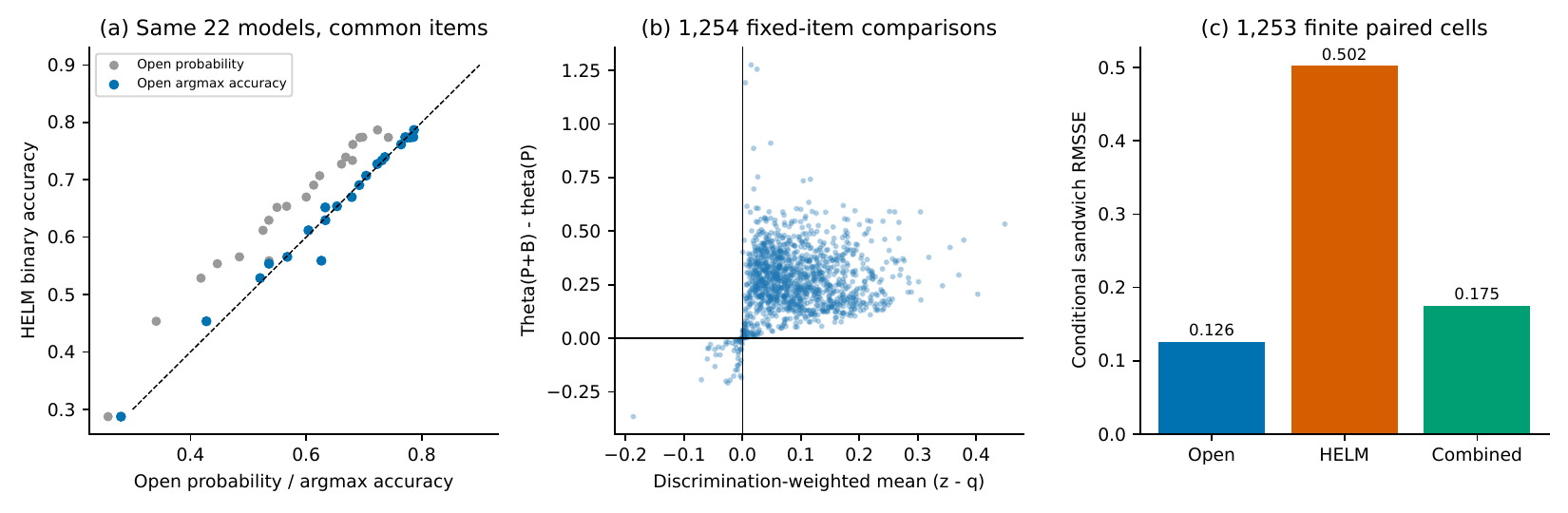}
\caption{
Overlap diagnostics on the fixed integrated item scale.
(a) compares the Open correct-option probability, Open argmax accuracy,
and HELM binary accuracy for the 22 overlapping models.
(b) plots model--subject ability shifts against the corresponding
discrimination-weighted channel difference.
(c) reports fixed-item analytical RMSSE for Open-only probability responses, HELM-only binary responses, and integrated responses.
}
\label{fig:empirical-overlap}
\end{figure}

To examine this difference, we converted the option probabilities in the Open LLM Leaderboard into binary responses by selecting the highest-probability option
\footnote{
We refer to the accuracy computed from these transformed binary responses as argmax accuracy.
}.
As shown in Figure~\ref{fig:empirical-overlap}(a), the resulting accuracy is much closer to aggregate HELM accuracy, whereas the mean correct-option probability in the Open LLM Leaderboard is systematically lower than HELM accuracy.
Hardening---selecting the most probable option instead of retaining its probability---is consistent with a substantial part of the observed ability drift. 
This agreement suggests that the distinction between probability responses and argmax correctness explains a substantial part of the aggregate discrepancy, although differences in prompts, deployments, and checkpoints may also contribute, as discussed in Appendix~\ref{sec:protocol}.

The relationship between the response definitions, ability drift, and uncertainty is summarized in Figure~\ref{fig:empirical-overlap}.

However, the agreement between the two sources is not exact at the individual model--subject level.
As illustrated in Figure~\ref{fig:empirical-overlap}(b), the size and even the direction of the remaining channel differences vary across models and subjects.

Figure~\ref{fig:empirical-overlap}(c) compares the RMSSE of ability estimates ($\theta$) for the 22 overlapping models using Open-only, HELM-only, and integrated response data, with the same fixed item parameters. 
The integrated condition yields lower ability RMSSE than HELM-only estimation, but higher RMSSE than Open-only estimation.

The simulations account for uncertainty from joint estimation of item and ability parameters, whereas this empirical comparison conditions on the fitted item parameters. 
The reported RMSSE therefore reflects conditional ability uncertainty rather than total uncertainty from joint calibration.

The analysis suggests that probability and binary responses can be placed in a common IRT framework, but statistical integration does not by itself guarantee measurement equivalence.
For practical applications, prompting, decoding, answer extraction, and response
definitions should be aligned as closely as possible across data sources before the channels are combined.

\subsection{Subject-level item characteristics}
Item-level benchmarking research has used IRT to assess measurement quality across 11 benchmarks and 41,871 items \citep{zhou2026benchmarks}. 
Here we examine heterogeneity in item discrimination and difficulty across MMLU subjects on their respective fitted scales.
\begin{table}[tbp]
\centering\small
\caption{The three subjects with the highest and the three with the lowest mean item discrimination $a$, including zero slopes ($a =0$). 
Item parameters use the corresponding subject-specific ability scale. 
Difficulty $b$ is summarized by the median over positive-slope items.}
\label{tab:empirical-domains}
\begin{tabular}{p{.40\linewidth}rrrr}\toprule
Subject & Items & Mean $\hat a$ & $a=0$ & Median $\hat b$ \\
\midrule High school government and politics & 193 & 1.933 & 5 & -0.944 \\
Miscellaneous & 783 & 1.882 & 97 & -0.934 \\
High school psychology & 545 & 1.870 & 62 & -0.902 \\
College mathematics & 100 & 0.241 & 58 & 1.189 \\
Abstract algebra & 100 & 0.174 & 46 & 2.106 \\
High school mathematics & 270 & 0.158 & 121 & 2.328 \\
\bottomrule

\end{tabular}
\end{table}

Table~\ref{tab:empirical-domains} lists the subjects selected by mean item discrimination
\footnote{
Table~\ref{tab:all-domains} in Appendix~\ref{sec:all-subjects} reports median difficulty for positive-slope items in all 57 subjects.
}
, using the same selection rule throughout the examples. 
High school government and politics, miscellaneous, and high school psychology have the largest mean discrimination values; college mathematics, abstract algebra, and high school mathematics have the smallest. 
Near-zero discrimination makes $b=-\beta/c$ an unstable extrapolated ratio, so its mean or RMSSE can be dominated by a single poorly identified item. 

These summaries describe the relationship between the items and the analyzed model population. 
They suggest that the selected mathematics subjects are relatively difficult and provide weaker separation along the fitted one-dimensional ability axis. 

\subsection{Dimensionality diagnostics with principal component analysis}
\label{subsec:dimensionality}
\begin{figure}[!t]
\centering\includegraphics[width=\linewidth]{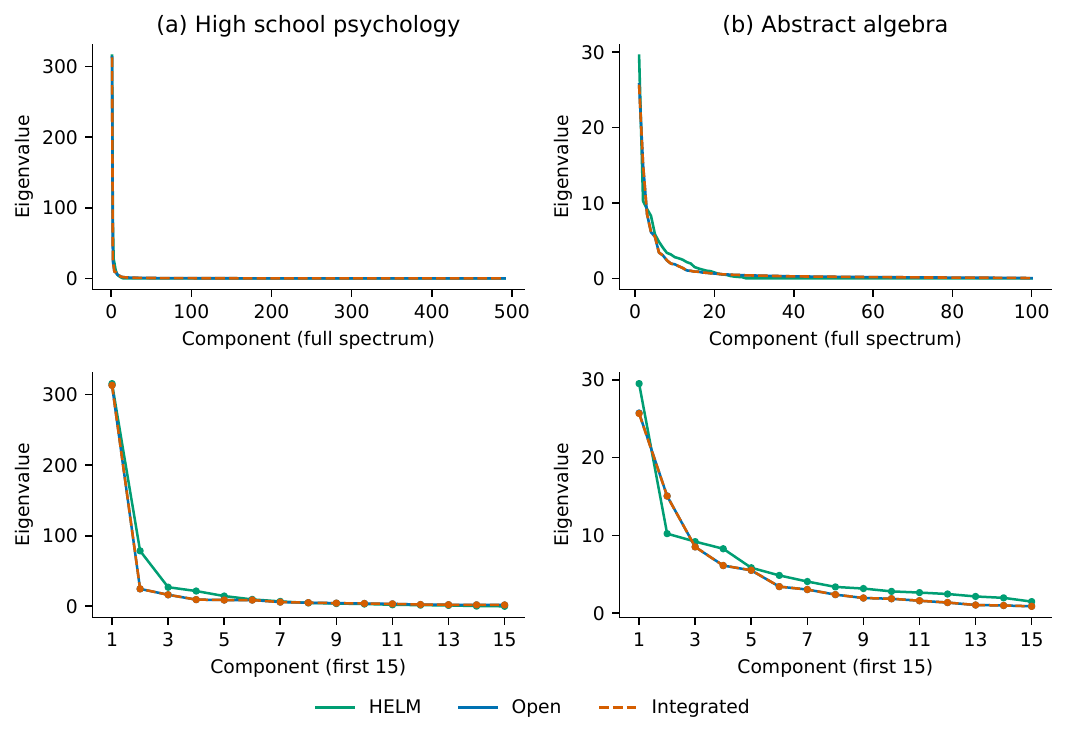}
\caption{Representative item-correlation scree plots obtained from principal component analysis of the estimated item correlation matrices: high school psychology (smallest integrated one-factor SRMR among the six examples in Tables~\ref{tab:empirical-domains} and~\ref{tab:dim-examples}) and abstract algebra (largest within that set). 
The upper panels show full spectra, and the lower panels show the largest 15 components. Within each subject, HELM-only, Open-only, and integrated analyses use the same retained item set. 
Appendix~\ref{sec:mixed-dim-theory} describes the mixed-format method, and Appendix~\ref{sec:mixed-dim-mmlu} presents the MMLU diagnostics.
}
\label{fig:empirical-dim}
\par\medskip
{\small
\pairedtablecaption{Integrated dimensionality diagnostics for the same six subjects and in the same order as Table~\ref{tab:empirical-domains}. 
$J$ is the common retained item count for the three correlation analyses; it can be smaller than the IRT item count in Table~\ref{tab:empirical-domains} because the correlation diagnostics apply additional item screening. 
$\lambda_1$ and $\lambda_2$ are the largest and second-largest eigenvalues of the item correlation matrix $\widehat{R}$ with positive-semidefinite~(PSD) correction.
SRMR is calculated for a separate one-factor fit to the uncorrected pairwise matrix $\widetilde{R}$.
Appendices~\ref{sec:mixed-dim-theory} and~\ref{sec:mixed-dim-mmlu} describe the method and the MMLU analysis, respectively.
}
\label{tab:dim-examples}
\begin{tabular}{p{.43\linewidth}rrrr}\toprule
Subject & $J$ & $\lambda_1/J$ & $\lambda_1/\lambda_2$ & SRMR\\\midrule
\input{tables/dimensionality_examples.tex}
\end{tabular}
}
\end{figure}

Figure~\ref{fig:empirical-dim} compares principal-component scree plots for high school psychology and abstract algebra using matched item sets in the HELM-only, Open-only, and integrated analyses.
\footnote{
Constructing a principal-component diagnostic after integrating probability and binary responses requires assumptions beyond their common IRT mean. 
Appendix~\ref{sec:mixed-dim-theory} formulates the mixed latent-correlation approach and examines it by simulation, while Appendix~\ref{sec:mixed-dim-mmlu} reports the MMLU checks and all 57 subject-level comparisons.
}

The enlarged panels show a sharper decline after the first eigenvalue in high school psychology than in abstract algebra, suggesting differences in the relative dominance of the first principal component. 
The integrated spectra are nearly identical to the Open-only spectra, consistent with the much larger Open sample. The common latent-correlation assumption remains unverified for these sources.

To quantify these differences, we use the proportion of variance explained by the first principal component ($\lambda_1/J$), the ratio of the first two eigenvalues ($\lambda_1/\lambda_2$), and the standardized root mean square residual (SRMR) from a separate one-factor correlation fit, where $\lambda_1$ and $\lambda_2$ are the largest and second-largest eigenvalues. 
Table~\ref{tab:dim-examples} summarizes these measures for the same six subjects, in the same order as Table~\ref{tab:empirical-domains}. 
Among these six subjects, high school psychology and abstract algebra have the smallest and largest integrated SRMR, respectively, motivating their selection as illustrative examples in Figure~\ref{fig:empirical-dim}. 

The scree plots compare the dominance of the leading component across subjects. The residuals separately indicate how closely a one-factor model approximates the estimated correlations. 
Both diagnostics are exploratory and identify subjects for which the unidimensional IRT assumption warrants closer assessment. 
%Appendix~\ref{sec:mixed-dim-mmlu} further examines the effects of sparse binary data, matrix correction, and the common-correlation assumption.

\section{Conclusion}
\paragraph{Contributions and limitations}\mbox{}

We integrate binary correctness and fractional correct-option probabilities in a common 2PL estimating criterion, including a shared ability for overlapping respondents. 
This use of established fractional-response QMLE includes binary-only models alongside models providing probabilities without discarding the latter information.
Broader measurement coverage supports comparisons among models important to an evaluation, even when their inclusion does not improve item-calibration precision.
Meaningful common-scale comparability still depends on appropriate measurement assumptions and aligned evaluation conditions.

The simulation results show that, when probability and binary responses share the same conditional mean, adding probability respondents improves item calibration more efficiently than adding binary respondents. 
However, adding noisy binary respondents to a probability sample can worsen both estimation error and reported uncertainty. 
Ability estimates for the original respondents change little in either direction.
Thus, the compatibility of probability and binary responses within the fractional-response QMLE framework does not guarantee that their equal-weight integration improves statistical efficiency.

The MMLU application reveals family-associated response patterns and subject-level differences in item characteristics. 
It also shows systematic ability shifts between Open and HELM records, whose evaluation protocols differ. 
These observations limit a common-mean interpretation of the integrated estimates. The dimensionality diagnostics are similarly exploratory: their dependence on the much larger probability sample and sparse binary data prevents a general claim that one dimension is adequate. 
The framework broadens the range of LLMs that can be analyzed on a common scale, while demonstrating that broader coverage, statistical efficiency, and measurement validity are distinct considerations.

\paragraph{Future directions}\mbox{}

Future research could explore response weighting schemes that account for differences in response variability, along with more controlled evaluation protocols to support the common-mean assumption.

Given the potential limitations of unidimensionality discussed in Section~\ref{subsec:dimensionality}, another direction is to extend the proposed framework to multidimensional IRT. 
Such an extension would require appropriate treatment of scale identification, cross-source measurement invariance, and joint uncertainty estimation.

From an application perspective, human--LLM joint calibration represents another promising direction, building on previous research comparing human and AI responses \citep{lalor2016,gor2024caimira,liu2025respondents,ebara2026}. 
Probability responses from LLMs could improve item calibration when human response samples are limited, provided that both populations can be placed on a defensible common scale. 
Investigating this possibility would require careful assessment of construct comparability, differential item functioning, and dependence among related models.

\section*{Acknowledgements}
\label{sec:Acknowledgements}
This work was supported by the Japan Science and Technology Agency (JST) under CREST Grant Number JPMJCR22D2.

\bibliographystyle{plainnat}
\bibliography{references}

\clearpage

\appendix
\section*{Appendix}
\addcontentsline{toc}{section}{Appendix}
\section{Related work}
\label{appendix: related work}
This review situates the contribution among response models, LLM evaluation methods, and Human--AI measurement. It distinguishes integration of response formats from adaptive testing, machine-assisted item calibration, and scoring models on a human reference scale.

\paragraph{IRT foundations and response formats}\mbox{}\

IRT separates properties of items from the latent abilities of respondents through a conditional response model \citep{lord1980}.
A widely used specification is the two-parameter logistic (2PL) model, which represents item difficulty and discrimination while retaining computational tractability and interpretability.
Because the latent ability scale has no intrinsic origin or unit, its location and scale must be fixed for model identification, typically through a normalization or standardization constraint.
An ability estimate is not automatically a population-invariant measure: dimensionality, local independence, item fit, and invariance across respondent groups remain substantive conditions \citep{reise1993}.

Extensions beyond correct/incorrect scoring follow more than one route. 
The nominal response model of \citet{bock1972} assigns category-specific parameters to the alternatives of an item, allowing the identities of incorrect responses to contribute information about ability. 
It models unordered answer categories. 
\citet{samejima1973}, in contrast, develops a response model for genuinely continuous observations.
This distinction matters for LLM evaluation: a selected answer category, an entire vector of option probabilities, and a scalar probability assigned to the correct option are different observations. 

\paragraph{LLM benchmarking and the evaluation of generated responses}\mbox{}\

LLM benchmarks differ not only in the capabilities they target, but also in the form of model responses and the mechanisms used to score them.
Closed-ended benchmarks typically define a fixed set of candidate answers and permit direct, objective scoring, whereas open-ended benchmarks require evaluation of generated responses.
The present study focuses on closed-ended multiple-choice evaluation because it can provide both binary correctness and, when available, probabilities assigned to the candidate options.

MMLU, for example, evaluates multiple-choice knowledge and reasoning across 57 subjects \citep{hendrycks2021}. 
HELM places accuracy within a broader evaluation framework covering multiple scenarios and metrics \citep{liang2023}. 
Such infrastructure improves reproducibility, but a benchmark score still depends on the selected tasks, prompts, answer extraction, and model population. 
Even changing answer-option order can alter multiple-choice performance \citep{pezeshkpour2024}. A common leaderboard does not by itself establish a common latent construct or equivalent evaluation protocols.

Confidence elicitation is a related but distinct measurement problem. \citet{xiong2024confidence} examine verbalized confidence, repeated response sampling, and aggregation based on response consistency, finding substantial calibration limitations. Such elicited confidence scores are not the API-exposed correct-option probabilities used here; they cannot be substituted without checking their interpretation and conditional-mean relationship.

Open-ended evaluation requires scoring generated responses, often with LLM-based evaluators such as MT-Bench and Chatbot Arena \citep{zheng2023}. Judge-based scores can exhibit position and verbosity biases \citep{li2025judge}. Generated responses can also be scored objectively against verifiable constraints, as in IFEval \citep{zhou2023}. These scores have different interpretations from binary correctness and correct-option probabilities and fall outside the present framework.

\paragraph{IRT in LLM evaluation}\mbox{}\

IRT has been introduced into LLM evaluation from several perspectives.
In the context of the reliability and diagnostic analysis of language-model and natural language processing leaderboards, \citet{rodriguez2021} use a Bayesian item-response approach to examine leaderboard reliability, informative examples, and annotation errors.
A second line of work focuses on measurement efficiency.
\citet{zhuang2025} advocate computerized adaptive testing for LLM evaluation, with questions selected adaptively according to model performance. 
ATLAS uses Fisher-information-guided adaptive item selection and reports substantial item savings while retaining ability-estimation precision \citep{li2026atlas}. \citet{zheng2026medicalcat} also study IRT-based adaptive evaluation of medical knowledge using simulations and 38 LLMs. Their focus is adaptive item selection; our focus is joint calibration from binary and probability records.
These studies show why item informativeness matters beyond the total number of correct answers.

Several later works make IRT-based LLM evaluation more practical. 
tinyBenchmarks uses IRT-related representations and selected examples to estimate full-benchmark performance with evaluations on fewer items \citep{polo2024}. 
Metabench calibrates item-response models on more than 5,000 LLM records, selects informative items from six benchmarks, and studies the structure of benchmark-specific abilities \citep{kipnis2025}. Using an IRT-based benchmark quality model, \citet{zhou2026benchmarks} examine measurement quality across 11 benchmarks and 41,871 items.
\citet{land2026audit} extend item-level diagnosis to benchmark auditing, using IRT to identify likely labeling errors and ambiguous items.
The framework of \citet{liao2025} supports binary and continuous evaluation metrics with a factorized ability structure across metrics or benchmarks. 
Building on the nominal response model \citep{bock1972}, \citet{fei2026} model the full distribution over answer options in multiple-choice questions and estimate option-level characteristics together with LLM abilities and additional response tendencies.
\citet{truong2026} introduce item response scaling laws and instantiate them with Beta-IRT, using token probabilities and empirical pass rates as richer measurements than binary outcomes. This provides a close precedent for probability-valued IRT measurement. Our focus is instead joint estimation of heterogeneous binary and fractional channels through a common 2PL mean and fractional-response QMLE, including partially overlapping respondents with a shared ability, without specifying a probability-response density.

A recent preprint by \citet{jiang2026} examines how small model samples, large item banks, and non-normal ability distributions affect IRT estimation for AI benchmarks. Its simulations motivate checking item recovery and numerical feasibility in the respondent--item regime used for AI evaluation.

\paragraph{Human--AI comparisons and machine-assisted item calibration}\mbox{}\

Human--AI comparison using IRT is an established research direction. 
\citet{lalor2016} calibrate a textual interpretation scale with human responses and use it to compare natural language processing (NLP) systems with a human population. 
Moving beyond human-only calibration, \citet{lalor2019} estimate item properties from artificial crowds of neural models while also identifying discrepancies between human and machine difficulty. 
More recently, the model of \citet{gor2024caimira} jointly learns multidimensional agent skills and content-dependent item characteristics from binary responses of 155 humans and approximately 70 AI systems. 
These studies include both human-referenced scoring and joint or machine-based calibration.

In education, \citet{liu2025respondents} compare item calibration from six LLMs with human calibration and evaluate augmentation of a small human sample. 
Their results show not only potential benefits but also narrow LLM proficiency distributions and sensitivity to the augmentation strategy.
In medical ethics, \citet{ebara2026} compare four LLMs with three physicians on 50 branching clinical scenarios, using 2PL models for binary scenario scores and graded-response models for ordinal scores. 
Their methods describe both physician and LLM responses as inputs to item estimation, with repeated LLM runs treated as simulated respondents; this is not a calibrated item bank only for humans. 
These scored trajectories are distinct from observed correct-option probabilities. 
The small physician reference sample also limits population-level interpretation.

\paragraph{Handling mixed response formats and fractional-response QMLE}\mbox{}\

Combining heterogeneous response formats within a common latent-variable model generally requires more than sharing a latent ability parameter.
The response types may differ in support, conditional variance, and distributional form, and their observations must nevertheless be linked to a common underlying quantity if they are to be pooled on the same measurement scale.
A conventional approach is to specify a separate observation model for each response type while connecting them through shared latent parameters, as in generalized latent-trait models for mixed exponential-family responses \citep{moustaki2000}.
For non-binary responses, however, this typically requires additional assumptions about their conditional distribution and variance.

The probability responses considered here have a particular structure: they are bounded in $[0,1]$ and represent the probability assigned to the correct option.
This makes fractional-response quasi-maximum likelihood a natural choice for the present setting. More generally, quasi-likelihood separates estimation based on a mean--variance relationship from specification of a full response distribution \citep{wedderburn1974}.
\citet{papke1996} develop regression methods for fractional outcomes in $[0,1]$ using a Bernoulli-form quasi-likelihood, including observations at zero and one.
The method specifies the expected fractional response without assigning it a Bernoulli distribution.

For a logistic expectation $\mu(\eta)$, the criterion contribution
$q\log\mu+(1-q)\log(1-\mu)$
has score $q-\mu$ with respect to $\eta$.
The same score form applies to a binary response when $q\in\{0,1\}$ and to a fractional response when $q\in[0,1]$, provided that both share the conditional expectation $\mu$.
A binomial count similarly contributes $K-R\mu$.
This common score structure provides a natural way to combine binary and probability responses within the same 2PL estimating criterion.

The distinction between expectation and variance is essential.
A common estimating equation does not imply equal conditional variances, independence between channels, or efficiency of unit weights.
The inverse curvature of the working criterion alone therefore need not provide a valid estimate of the sampling variance.
Misspecified-likelihood covariance theory \citep{white1982} motivates the score-based covariance calculation below, while complete refits under pairs resampling provide a complementary assessment \citep{efron1979}.
These classical results concern settings whose regularity conditions must be checked; they are not, by themselves, consistency or normality theorems for a jointly estimated IRT model whose number of nuisance parameters grows with $N$ and $J$ \citep{chen2019}.
Consequently, we assess parameter recovery and uncertainty scaling in controlled finite-sample experiments.

\section{Details of simulations}\label{sec:detailed design}
This appendix describes the initialization, numerical estimation, and response generator used in the simulations in Section~\ref{sec:design}.

\subsection{Simulation initialization and numerical estimation}
\label{subsec: initialization}
Let $Y,W$ be the aggregate responses and repeat counts in Eq.~\eqref{eq:aggregate}. Define
\[
 u_i=\operatorname{logit}\!\left[\operatorname{clip}
 \left\{\frac{\sum_jY_{ij}+0.5}{\sum_jW_{ij}+1},0.001,0.999\right\}\right].
\]
The simulations are initialized with $\theta_i=(u_i-\bar u)/\max\{\operatorname{SD}(u),0.1\}$, with SD calculated using the divisor $N$, $c_j=1$, and
\[
 \beta_j=\operatorname{logit}\!\left[\operatorname{clip}
 \left\{\frac{\sum_iY_{ij}+0.5}{\sum_iW_{ij}+1},0.001,0.999\right\}\right],
\]
followed by the specified identification transformation. Here SD denotes standard deviation and $\operatorname{clip}$ restricts a value to the stated interval. The pseudo-counts initialize the optimizer without modifying the observations. Each block is updated using damped Newton steps toward the minimizers in Algorithm~\ref{alg:fit}; the implementation does not require exact block minimization at every iteration. Simulation convergence is assessed from changes in abilities, log slopes, and intercepts, and analytical covariance is reported only when the constrained curvature checks pass. Empirical refits use archived warm-start arrays rather than newly generated response proportions. Numerical tolerances, iteration limits, and warm-start specifications are documented in the accompanying computational specification.

\subsection{Details of data generation}
For both response formats, we generate $a_j\sim\mathrm{Uniform}(0.5,2)$, $b_j\sim\mathrm{Uniform}(-2,2)$, and $\theta_i\sim N(0,1)$ independently. 
Each available response channel has five repeats. 
Binary responses are generated as $K_{ij}\sim\mathrm{Binomial}(5,\mu_{ij})$, where $\mu_{ij}$ is the common conditional mean specified by the 2PL model.
For probability responses, we must specify how to simulate variability in LLM probability measurements.

\paragraph{Conditional variance of probability and binary responses}\mbox{}\\
Under the common conditional-mean assumption in Eq. \eqref{eq:restriction}, probability responses satisfy $E(Q_{ijr}\mid\theta_i,a_j,b_j)=\mu_{ij}$ and $0\leq Q_{ijr}\leq1$. 
Because $Q_{ijr}^2\leq Q_{ijr}$, their conditional variance is bounded by

\[
\begin{aligned}
\operatorname{Var}(Q_{ijr}\mid\theta_i,a_j,b_j)
&=E(Q_{ijr}^2\mid\theta_i,a_j,b_j)-\mu_{ij}^2\
&\leq\mu_{ij}-\mu_{ij}^2\
%&=\mu_{ij}(1-\mu_{ij})
&=\operatorname{Var}(X_{ijr}\mid\theta_i,a_j,b_j),
\end{aligned}
\]
where the binary response $X_{ijr}\sim\operatorname{Bernoulli}(\mu_{ij})$. 
%The inequality is strict when the probability response takes values strictly between zero and one with positive probability. 
Thus, under the common conditional-mean assumption, probability responses have no greater conditional variance than binary correctness indicators, regardless of their specific distribution. 
However, this theoretical bound does not determine the magnitude of their variance.

\paragraph{Reported magnitude and behaviors of probability-response variability} 
Previous studies provide evidence on the variability of LLM probability outputs. 
\citet[Figure~16]{takayama2025} report an approximately quadratic relationship between mean response probability and the standard error of repeated token probability measurements from prompts requesting binary responses of ``yes'' or ``no'', with greater variability near intermediate probabilities and smaller variability toward the boundaries. 
Moreover, \citet[Figure 2]{Fu2026token_uncertainty} provide large-scale empirical evidence on the magnitude of token-probability fluctuations across repeated LLM inference runs. 
Their results show that the standard deviations are below 0.05 for almost all evaluated token probabilities across the tested models and GPU configurations, with most observations concentrated at substantially smaller values. 

Although the two studies report different variability measures and do not directly estimate the conditional residual variance of a 2PL model, their findings provide empirical guidance on the magnitude and probability-dependent shape of LLM probability fluctuations.

\paragraph{Specification of probability data generation}
Guided by these theoretical and empirical considerations, we adopt the following heuristic conditional variance function for probability responses:

\begin{equation}
  v(\mu)=\min\left\{
    \left[0.01+0.2\sqrt{\mu(1-\mu)}\right]^2,
    0.5\mu(1-\mu)
  \right\}.
\end{equation}

This function assigns smaller variance near the boundaries and larger variance at intermediate response probabilities. 
The second term ensures that the specified variance remains below the Bernoulli variance and is feasible for a beta distribution. 
%For example, at $\mu=0.5$, the specified probability-response variance is $0.0121$, compared with the Bernoulli variance of $0.25$. 
%The numerical coefficients are simulation choices informed by the qualitative empirical patterns, rather than estimates fitted to the studies cited above.

We generate probability observations from a beta distribution parameterized to have conditional mean $\mu_{ij}$ and variance $v(\mu_{ij})$. 
This ensures that the simulated probability $Q_{ijr}$ satisfies $0\leq Q_{ijr}\leq 1$ while preserving the common conditional mean, without clipping. 
%The beta distribution is used only for data generation; neither its distributional form nor its variance function is assumed in the fractional-response QMLE. 
Once the true parameters are generated, response observations are generated independently across repeats, channels, and model--item cells.

The simulation therefore evaluates estimation performance under a probability-response variance structure motivated by theoretical bounds and empirical observations of LLM probability variability. 
Although the variance ordering $\operatorname{Var}(Q_{ijr}\mid\theta_i,a_j,b_j) \leq \operatorname{Var}(X_{ijr}\mid\theta_i,a_j,b_j)$ is general under the common-mean assumption, the magnitude of the resulting precision advantage depends on the specified variance, measurement design, and estimation procedure. 

\section{Empirical estimation and numerical diagnostics in the evaluation with MMLU}\label{sec:empirical-computation}
The MMLU analysis in Section~\ref{sec:empirical} uses a weakly penalized estimator to stabilize poorly identified directions.
The empirical parameterization is $\eta_{ij}=t_i c_j+\beta_j$, $c_j\geq0$, with $c_j=1.702a_j$ on the reporting scale. 
Here $c$ is a slope, not a guessing parameter. 
The penalized criterion minimized in the empirical analysis is
\begin{equation}
 \sum_{ij}\{w_{ij}\log(1+e^{\eta_{ij}})-y_{ij}\eta_{ij}\}
 +\frac{\lambda}{2}\left[\sum_i t_i^2+
 \frac{N}{J}\sum_j(c_j^2+\beta_j^2)\right].
\end{equation}
The penalty is applied to the summed loss, not its mean, and is not a penalty on $\log a$ or $b$. 
The primary coefficient is $\lambda=10^{-6}$. 
The fitted abilities are standardized using eligible Open LLM Leaderboard respondents as the reference group. The corresponding transformations of item parameters and analytical SEs preserve the fitted response probabilities.

Items with zero estimated discrimination are retained in the model as constant-response items, but their difficulty parameters are undefined. Respondents with insufficient item coverage or information for stable ability estimation are excluded according to prespecified criteria. The primary analysis retains 372,721 model--subject ability estimates and all 14,042 items, including 3,697 zero-slope items. Convergence and ridge-sensitivity checks indicate numerical stability across the 57 subjects, although they do not establish global optimality or resolve weak identification.

Analytical SEs extend the constrained sandwich estimator in Section~\ref{sec:sandwich} to the penalized objective and standardized reporting scale. The calculation retains joint item--ability uncertainty but conditions on the fitted set of zero-slope items. Accordingly, the resulting SEs do not capture uncertainty associated with changes in boundary status.

Finally, item difficulty can become highly unstable when discrimination approaches zero, because $b=-\beta/c$. Even when the fitted response curve remains stable, a near-zero positive slope can yield an extreme difficulty estimate and SE. In the empirical analysis, a small number of such items dominate the raw difficulty RMSSE. We therefore interpret difficulty estimates cautiously near the zero-slope boundary.

\FloatBarrier
\section{Comparability of evaluation protocols of MMLU}\label{sec:protocol}

The empirical analysis includes 22 models represented in both the Open LLM Leaderboard and HELM datasets, as listed in Table~\ref{tab:overlap-matching}. 
These models share one ability parameter across response formats within each MMLU subject. 
However, matching model identifiers does not necessarily establish identical model checkpoints or evaluation conditions.

The two sources differ in how responses are obtained. 
Open LLM Leaderboard uses normalized candidate-option likelihoods, whereas HELM evaluates generated answers using temperature-zero decoding. 
Although both evaluations use five-shot prompting, differences in prompting, decoding, answer extraction, and model deployment may affect the resulting responses. In particular, deterministic correctness at temperature zero need not have the same conditional expectation as the correct-option probability obtained from candidate likelihoods.

Consequently, the common-mean assumption cannot be verified from these records alone. The integrated analysis therefore provides an empirical illustration of the proposed framework, rather than evidence that the two evaluation protocols yield equivalent measurements. A more direct assessment would require probability and binary responses obtained under matched prompts and sampling conditions from the same fixed model checkpoints.

\begin{table}[H]
\centering\small
\caption{Models shared by the Open LLM Leaderboard and HELM datasets.}
\label{tab:overlap-matching}
\begin{tabular}{@{}l@{\hspace{1.5em}}l@{}}
\toprule
Shared model & \\
\midrule
\texttt{01-ai/Yi-34B} & \texttt{google/gemma-7b} \\
\texttt{01-ai/Yi-6B} & \texttt{meta-llama/Llama-2-13b-hf} \\
\texttt{Qwen/Qwen1.5-110B-Chat} & \texttt{meta-llama/Llama-2-70b-hf} \\
\texttt{Qwen/Qwen1.5-14B} & \texttt{meta-llama/Llama-2-7b-hf} \\
\texttt{Qwen/Qwen1.5-32B} & \texttt{meta-llama/Meta-Llama-3-70B} \\
\texttt{Qwen/Qwen1.5-72B} & \texttt{meta-llama/Meta-Llama-3-8B} \\
\texttt{Qwen/Qwen1.5-7B} & \texttt{microsoft/Phi-3-medium-4k-instruct} \\
\texttt{allenai/OLMo-1.7-7B-hf} & \texttt{microsoft/phi-2} \\
\texttt{allenai/OLMo-7B-hf} & \texttt{mistralai/Mistral-7B-v0.1} \\
\texttt{databricks/dbrx-instruct} & \texttt{mistralai/Mixtral-8x22B-v0.1} \\
\texttt{deepseek-ai/deepseek-llm-67b-chat} & \texttt{mistralai/Mixtral-8x7B-v0.1} \\
\bottomrule

\end{tabular}
\end{table}

\section{All-subject empirical summaries in MMLU}\label{sec:all-subjects}
Table~\ref{tab:all-domains} extends the item summaries to all 57 subjects in MMLU, and Table~\ref{tab:all-overlap-se} compares overlap uncertainty on the fixed integrated item scale.
The tables separate subject-specific scales. 
Mean discrimination includes every retained item, including items with $a=0$. 
Difficulty summaries and item RMSSEs use only positive-slope items; undefined boundary values are recorded as missing. 

For the overlapping models, Open-only, HELM-only, and integrated ability estimates are obtained using the same fixed item parameters from the integrated fit. Their analytical SEs are calculated conditionally on these item parameters, without additional scale normalization. For integrated responses, probability and binary contributions from the same item are clustered together.

\clearpage
Figure~\ref{fig:empirical-overlap}(c) and Table~\ref{tab:all-overlap-se} summarize the resulting ability RMSSE, calculated across all finite model--subject estimates. One HELM-only estimate is excluded because of separation, leaving 1,253 comparable model--subject estimates. The joint SEs reported separately in the table additionally account for uncertainty in the estimated item parameters.
\begingroup\fontsize{8}{9.6}\selectfont\renewcommand{\arraystretch}{.95}\setlength{\LTpre}{0pt}\setlength{\LTpre}{6pt}
\begin{longtable}[t!]{p{.42\linewidth}rrrr}
\caption{All 57 subject-level item summaries.}\label{tab:all-domains}\\
\toprule Subject & Items & Mean $a$ & $a=0$ & Median $b$\\\midrule\endfirsthead
\toprule Subject & Items & Mean $a$ & $a=0$ & Median $b$\\\midrule\endhead
Abstract algebra & 100 & 0.174 & 46 & 2.106 \\
Anatomy & 135 & 1.056 & 39 & -0.538 \\
Astronomy & 152 & 1.131 & 14 & -0.511 \\
Business ethics & 100 & 0.754 & 21 & -0.559 \\
Clinical knowledge & 265 & 1.027 & 48 & -0.683 \\
College biology & 144 & 1.606 & 24 & -0.666 \\
College chemistry & 100 & 0.503 & 38 & 0.063 \\
College computer science & 100 & 0.601 & 32 & -0.156 \\
College mathematics & 100 & 0.241 & 58 & 1.189 \\
College medicine & 173 & 1.107 & 47 & -0.648 \\
College physics & 102 & 0.406 & 46 & 0.814 \\
Computer security & 100 & 1.396 & 21 & -0.894 \\
Conceptual physics & 235 & 0.639 & 64 & -0.315 \\
Econometrics & 114 & 0.592 & 47 & -0.073 \\
Electrical engineering & 145 & 0.537 & 38 & -0.347 \\
Elementary mathematics & 378 & 0.387 & 154 & 0.424 \\
Formal logic & 126 & 0.413 & 52 & 0.140 \\
Global facts & 100 & 0.260 & 54 & 0.548 \\
High school biology & 310 & 1.242 & 29 & -0.743 \\
High school chemistry & 203 & 0.684 & 85 & -0.246 \\
High school computer science & 100 & 1.578 & 23 & -0.628 \\
High school european history & 165 & 1.172 & 24 & -0.904 \\
High school geography & 198 & 1.539 & 24 & -0.876 \\
High school government and politics & 193 & 1.933 & 5 & -0.944 \\
High school macroeconomics & 390 & 1.079 & 105 & -0.581 \\
High school mathematics & 270 & 0.158 & 121 & 2.328 \\
High school microeconomics & 238 & 1.001 & 36 & -0.542 \\
High school physics & 151 & 0.310 & 77 & 0.461 \\
High school psychology & 545 & 1.870 & 62 & -0.902 \\
High school statistics & 216 & 0.480 & 62 & 0.067 \\
High school us history & 204 & 1.386 & 17 & -0.885 \\
High school world history & 237 & 1.655 & 33 & -0.892 \\
Human aging & 223 & 1.164 & 61 & -0.793 \\
Human sexuality & 131 & 1.251 & 22 & -0.703 \\
International law & 121 & 1.228 & 24 & -0.943 \\
Jurisprudence & 108 & 0.902 & 10 & -0.689 \\
Logical fallacies & 163 & 1.511 & 30 & -0.829 \\
Machine learning & 112 & 0.605 & 52 & -0.137 \\
Management & 103 & 1.303 & 15 & -0.960 \\
Marketing & 234 & 1.492 & 17 & -0.994 \\
Medical genetics & 100 & 1.392 & 16 & -0.733 \\
Miscellaneous & 783 & 1.882 & 97 & -0.934 \\
Moral disputes & 346 & 0.927 & 69 & -0.741 \\
Moral scenarios & 895 & 0.389 & 392 & 0.355 \\
Nutrition & 306 & 0.820 & 47 & -0.643 \\
Philosophy & 311 & 1.077 & 71 & -0.779 \\
Prehistory & 324 & 1.124 & 43 & -0.746 \\
Professional accounting & 282 & 0.490 & 112 & -0.006 \\
Professional law & 1534 & 0.421 & 667 & -0.188 \\
Professional medicine & 272 & 0.828 & 56 & -0.484 \\
Professional psychology & 612 & 1.040 & 150 & -0.681 \\
Public relations & 110 & 0.850 & 28 & -0.783 \\
Security studies & 245 & 1.050 & 49 & -0.761 \\
Sociology & 201 & 1.552 & 24 & -0.872 \\
Us foreign policy & 100 & 1.812 & 10 & -0.882 \\
Virology & 166 & 0.758 & 66 & -0.733 \\
World religions & 171 & 1.680 & 23 & -0.982 \\
\bottomrule\end{longtable}
\endgroup
%\begingroup\small
%For the fixed-item overlap comparison in Table~\ref{tab:all-overlap-se}, the scalar estimates are unpenalized roots with the integrated $a,b$ fixed; no new centering/scaling is applied. Let $s_j=c_j(y_j-w_j\mu_j)$ and $A=\sum_jw_jc_j^2\mu_j(1-\mu_j)$. The variance estimate with the integrated item parameters held fixed is
%\[
% \widehat{\Var}(\hat\theta)=\frac{J}{J-1}\,
% \frac{\sum_j(s_j-\bar s)^2}{A^2}.
%\]
%Combined scores sum the two channel contributions within item. One HELM-channel fit (Llama-3-70B in college biology) is separated on positive-slope items, leaving 1,253 paired finite cells. The global RMSSE is the square root of the mean of all squared cell SEs, not an average of subject RMSSEs. The last column is the separately estimated integrated active-face joint SE; the first three hold items fixed.

%\endgroup

%\clearpage

\clearpage
\begingroup\fontsize{8}{9.6}\selectfont\renewcommand{\arraystretch}{.95}\setlength{\LTpre}{0pt}
\begin{longtable}[t]{p{.35\linewidth}rrrrr}
\caption{Subject-specific and pooled ability RMSSE for overlapping models. P: Open probability; B: HELM binary; I: integrated. The first three columns hold items fixed; Joint I includes item estimation.}\label{tab:all-overlap-se}\\
\toprule Subject & $n$ & $P$ & $B$ & $I$ & Joint $I$\\\midrule\endfirsthead
\toprule Subject & $n$ & $P$ & $B$ & $I$ & Joint $I$\\\midrule\endhead
All (pooled) & 1253 & 0.126 & 0.502 & 0.175 & 0.239 \\
Abstract algebra & 22 & 0.186 & 0.512 & 0.280 & 0.391 \\
Anatomy & 22 & 0.129 & 0.353 & 0.184 & 0.208 \\
Astronomy & 22 & 0.243 & 1.150 & 0.346 & 0.418 \\
Business ethics & 22 & 0.195 & 0.507 & 0.264 & 0.481 \\
Clinical knowledge & 22 & 0.086 & 0.278 & 0.114 & 0.147 \\
College biology & 21 & 0.116 & 0.345 & 0.150 & 0.173 \\
College chemistry & 22 & 0.188 & 0.521 & 0.265 & 0.317 \\
College computer science & 22 & 0.124 & 0.598 & 0.182 & 0.210 \\
College mathematics & 22 & 0.124 & 0.385 & 0.183 & 0.232 \\
College medicine & 22 & 0.099 & 0.296 & 0.128 & 0.158 \\
College physics & 22 & 0.145 & 0.477 & 0.221 & 0.278 \\
Computer security & 22 & 0.094 & 0.348 & 0.128 & 0.190 \\
Conceptual physics & 22 & 0.112 & 0.358 & 0.154 & 0.212 \\
Econometrics & 22 & 0.104 & 0.305 & 0.152 & 0.181 \\
Electrical engineering & 22 & 0.130 & 0.499 & 0.186 & 0.228 \\
Elementary mathematics & 22 & 0.091 & 0.213 & 0.119 & 0.171 \\
Formal logic & 22 & 0.126 & 0.368 & 0.171 & 0.206 \\
Global facts & 22 & 0.154 & 0.408 & 0.222 & 0.273 \\
High school biology & 22 & 0.117 & 0.352 & 0.160 & 0.207 \\
High school chemistry & 22 & 0.100 & 0.230 & 0.128 & 0.152 \\
High school computer science & 22 & 0.089 & 0.329 & 0.119 & 0.161 \\
High school european history & 22 & 0.192 & 0.796 & 0.257 & 0.314 \\
High school geography & 22 & 0.130 & 0.492 & 0.169 & 0.227 \\
High school government and politics & 22 & 0.341 & 1.273 & 0.488 & 0.822 \\
High school macroeconomics & 22 & 0.067 & 0.176 & 0.087 & 0.106 \\
High school mathematics & 22 & 0.132 & 0.355 & 0.197 & 0.233 \\
High school microeconomics & 22 & 0.136 & 0.363 & 0.188 & 0.330 \\
High school physics & 22 & 0.112 & 0.315 & 0.155 & 0.209 \\
High school psychology & 22 & 0.066 & 0.206 & 0.089 & 0.110 \\
High school statistics & 22 & 0.106 & 0.316 & 0.148 & 0.206 \\
High school us history & 22 & 0.153 & 0.566 & 0.186 & 0.210 \\
High school world history & 22 & 0.136 & 0.418 & 0.152 & 0.191 \\
Human aging & 22 & 0.073 & 0.266 & 0.098 & 0.119 \\
Human sexuality & 22 & 0.094 & 0.312 & 0.129 & 0.150 \\
International law & 22 & 0.109 & 0.376 & 0.146 & 0.177 \\
Jurisprudence & 22 & 0.113 & 0.546 & 0.167 & 0.197 \\
Logical fallacies & 22 & 0.096 & 0.273 & 0.116 & 0.141 \\
Machine learning & 22 & 0.107 & 0.443 & 0.163 & 0.212 \\
Management & 22 & 0.138 & 0.543 & 0.171 & 0.210 \\
Marketing & 22 & 0.082 & 0.305 & 0.103 & 0.131 \\
Medical genetics & 22 & 0.203 & 1.904 & 0.303 & 0.433 \\
Miscellaneous & 22 & 0.104 & 0.206 & 0.119 & 0.144 \\
Moral disputes & 22 & 0.059 & 0.226 & 0.080 & 0.098 \\
Moral scenarios & 22 & 0.043 & 0.142 & 0.058 & 0.068 \\
Nutrition & 22 & 0.085 & 0.317 & 0.114 & 0.128 \\
Philosophy & 22 & 0.066 & 0.228 & 0.093 & 0.114 \\
Prehistory & 22 & 0.113 & 0.290 & 0.142 & 0.193 \\
Professional accounting & 22 & 0.088 & 0.244 & 0.114 & 0.136 \\
Professional law & 22 & 0.031 & 0.084 & 0.043 & 0.050 \\
Professional medicine & 22 & 0.078 & 0.228 & 0.106 & 0.131 \\
Professional psychology & 22 & 0.062 & 0.176 & 0.083 & 0.102 \\
Public relations & 22 & 0.118 & 0.384 & 0.162 & 0.196 \\
Security studies & 22 & 0.052 & 0.234 & 0.076 & 0.097 \\
Sociology & 22 & 0.074 & 0.339 & 0.103 & 0.130 \\
Us foreign policy & 22 & 0.095 & 0.738 & 0.142 & 0.215 \\
Virology & 22 & 0.074 & 0.253 & 0.102 & 0.131 \\
World religions & 22 & 0.084 & 0.246 & 0.116 & 0.136 \\
\bottomrule\end{longtable}
\endgroup
\clearpage

\section{Mixed-format correlation diagnostics for dimensionality assessment}\label{sec:mixed-dim-theory}

This appendix describes an exploratory principal component analysis (PCA) of item responses observed as binary correctness indicators or continuous correct-option probabilities. 
The 2PL IRT model used in this paper assumes a single latent ability within each subject.
Therefore, it is important to assess the plausibility of the unidimensionality assumption in empirical applications. 
Examining the eigenvalue structure of the item correlation matrix provides a practical diagnostic for assessing whether a dominant component is present in the response data.
Although such analyses commonly use tetrachoric correlations for binary responses, extending this diagnostic of unidimensionality to the integrated analyses of continuous probability responses and binary responses requires a new framework that can be interpreted on a comparable latent scale.

We use a latent Gaussian-copula framework to construct a common item correlation matrix from the two response formats. 
We first review the tetrachoric approach for binary data, then introduce the extension to probability responses and the additional assumption required to combine the formats.
A simulation illustrates how the item-correlation spectrum can change across formats even when they share the same 2PL conditional mean. 
Although this correlation-based diagnostic is complementary to our fractional-response IRT framework, it relies on assumptions distinct from the common conditional-mean specification used for IRT estimation.

\subsection{Dimensionality assessment and tetrachoric correlations for binary responses}

Our 2PL IRT model specifies the expected response of LLM $i$ to item $j$ in Eq.~\eqref{eq:mean}.
Let $R=(\rho_{jk})$ denote an item correlation matrix with ordered eigenvalues $\lambda_1\geq\lambda_2\geq\cdots\geq\lambda_J$. 
A relatively large first eigenvalue, followed by a sharp decline in a scree plot, suggests that one component dominates the correlation structure. 
%This is an exploratory diagnostic rather than a test of all assumptions of a unidimensional IRT model.

For binary responses, tetrachoric correlation estimates the association between two underlying continuous latent variables. 
Specifically, we assume that each binary response $Y_j$ is generated by thresholding a latent variable $Z_j$,
\begin{equation}
Y_j=\mathbf{1}(Z_j>\tau_j),
\end{equation}
and that the latent variables for items $j$ and $k$ follow a bivariate standard-normal distribution with correlation $\rho_{jk}$, whose joint density is given by
\begin{equation}
\phi_2(z_j,z_k;\rho_{jk})
=
\frac{1}{2\pi\sqrt{1-\rho_{jk}^{2}}}
\exp\left(
-\frac{z_j^2-2\rho_{jk}z_jz_k+z_k^2}
{2(1-\rho_{jk}^{2})}
\right),
\qquad -1<\rho_{jk}<1.
\end{equation}

Let $\Phi$ denote the standard-normal cumulative distribution function,
\begin{equation}
\Phi(x)=
\int_{-\infty}^{x}
\frac{1}{\sqrt{2\pi}}
\exp\left(-\frac{z^2}{2}\right)\,dz,
\end{equation}
and let $\Phi_2$ denote the bivariate standard-normal cumulative distribution function with correlation $\rho$,
\begin{equation}
\Phi_2(x,y;\rho)=
\int_{-\infty}^{x}\int_{-\infty}^{y}
\phi_2(z_1,z_2;\rho)\,dz_2\,dz_1.
\end{equation}

Under this assumption, the probabilities of the four binary response patterns $\pi_{uv,jk}(\rho)$ are determined by the thresholds and the latent correlation:
\begin{equation}
\begin{aligned}
\pi_{00,jk}(\rho)
&=\Phi_2(\tau_j,\tau_k;\rho),\\
\pi_{01,jk}(\rho)
&=\Phi(\tau_j)-\Phi_2(\tau_j,\tau_k;\rho),\\
\pi_{10,jk}(\rho)
&=\Phi(\tau_k)-\Phi_2(\tau_j,\tau_k;\rho),\\
\pi_{11,jk}(\rho)
&=1-\Phi(\tau_j)-\Phi(\tau_k)
+\Phi_2(\tau_j,\tau_k;\rho).
\end{aligned}
\end{equation}

The true threshold $\tau_j$ is unknown and estimated from the marginal frequency of incorrect responses to that item as
\begin{equation}
 \widehat\tau_j=\Phi^{-1}\!\left(\frac{n_{0j}+1/2}{N_B+1}\right).
\end{equation}
Here, $n_{0j}$ denotes the number of incorrect responses on item $j$ among $N_B$ binary respondents.
The half-count correction prevents a marginal probability of exactly zero or one; it does not add a half-count to every cell of a $2\times2$ table.
Constant-response items are excluded because their associations are not estimable.
The estimated thresholds $\hat{\tau}_j$ and $\hat{\tau}_k$ are then substituted into the response-pattern probabilities $\pi_{uv,jk}(\rho)$ when estimating the latent correlation $\rho_{jk}$.

We then estimate $\rho_{jk}$ by choosing the value that best reproduces the observed frequencies of the four response patterns. Let $n_{uv,jk}$ denote the number of observed pairs with $(Y_j,Y_k)=(u,v)$. The two-step tetrachoric estimator \citep{olsson1979} maximizes
\begin{equation}
\hat{\rho}_{jk}
=
\underset{-1<\rho<1}{\arg\max}
\sum_{u,v\in\{0,1\}}
n_{uv,jk}\log\pi_{uv,jk}(\rho).
\end{equation}
Equivalently, this minimizes the Kullback--Leibler divergence between the empirical distribution of the four response patterns and their probabilities under the latent Gaussian model, treating the estimated thresholds as fixed. 
Repeating this estimation for each item pair yields a latent item correlation matrix $\widetilde R=(\widehat\rho_{jk})$.

However, $\widetilde R$ may not be positive semidefinite because its entries were fitted separately. 
Therefore, the nearest positive-semidefinite~(PSD) correlation matrix $\widehat R$ is constructed as
\begin{equation}\label{eq: PSD}
 \widehat R
 =\underset{R\succeq0,\;\operatorname{diag}(R)=\mathbf{1}}{\arg\min}
 \|R-\widetilde R\|_F^2,
\end{equation}
using nearest-correlation methods \citep{higham2002}. 
Here $\|\cdot\|_F$ denotes the Frobenius norm. 
%To assess the effect of this step, we retain the original matrix, its negative eigenvalues, and the relative correction $\|\widehat R-\widetilde R\|_F/\|\widetilde R\|_F$.

\subsection{Introducing Gaussian-copula framework to construct a common item correlation matrix}

In addition to binary indicators, our data contain correct-option probabilities $Q_{ij}\in[0,1]$. 
Pearson correlations between observed probabilities can be computed directly, but they measure linear associations between the probability values themselves. 
Tetrachoric correlations, by contrast, concern latent continuous variables. 
To extend this perspective to probability responses while retaining a common basis for comparison, we use a Gaussian-copula representation to place the two response formats on a common latent-correlation scale for dimensionality assessment.

A copula separates the marginal distributions of individual variables from their dependence structure. 
For a continuously distributed probability response $Q_j$ with cumulative distribution function $F_j$, the probability-integral transform followed by a standard-normal quantile transform gives
\begin{equation}
 Z_j=\Phi^{-1}\!\{F_j(Q_j)\},
\end{equation}
The transformed variable has a standard-normal marginal distribution when $F_j$ is continuous. 
A Gaussian copula further assumes a multivariate-normal dependence structure among these transformed variables; it does \emph{not} require the original probabilities $Q_j$ to be normally distributed. 
This use of marginal normalizing transformations is central to semiparametric Gaussian-copula models \citep{liu2009nonparanormal}. 
Latent Gaussian-copula models also provide a framework for combining continuous and binary measurements \citep{fan2014latent}.

However, because the marginal distributions $F_j$ are unknown, we approximate this transformation using ranks within the probability-response source. 
Let $N_P$ be the number of probability respondents and $r_{ij}=\operatorname{rank}_{P}(Q_{ij})$ be the ascending rank of response $Q_{ij}$ among those respondents for item $j$. 
We define the \emph{normal score}
\begin{equation}
 Z^P_{ij}=\Phi^{-1}\!\left(\frac{r_{ij}-1/2}{N_P}\right).
\end{equation}
%The plotting position $(r_{ij}-1/2)/N_P$ assigns the midpoint of each empirical cumulative-probability interval to its observation; in particular, it prevents the largest rank from mapping to $\Phi^{-1}(1)=+\infty$. 
We use average ranks for ties and subsequently center and scale each column to empirical mean zero and variance one. 
Normal scores preserve the order of probability responses but not the distances between their original values. 
Thus, their correlations target latent dependence rather than Pearson correlation on the original probability scale. 
For the present diagnostic, these rank-based associations provide a complementary view of common item dependence, although their interpretation differs from that of tetrachoric correlations.

To combine the formats, we make the additional \emph{working assumption} that, for each item pair $(j,k)$, the latent variables in the probability and binary sources have a common correlation:
\begin{equation}
 \operatorname{Corr}(Z^P_j,Z^P_k)
 =\operatorname{Corr}(Z^B_j,Z^B_k)
 =\rho_{jk}.
\end{equation}
The two sources may have different marginal response distributions or thresholds; only their latent item correlations are constrained to be common. 
The observation pattern here differs from the usual mixed-variable setting: the same items are recorded in different formats for different groups of LLMs, rather than different variables having fixed formats within each respondent. 
Accordingly, this common-correlation restriction is an additional assumption of our diagnostic, not a consequence of the common 2PL conditional mean used in the fractional-response IRT model.

\subsection{Joint pairwise correlation estimation and principal component analysis}
For probability responses, the transformed values $Z^P_{ij}$ and $Z^P_{ik}$ are observed as continuous normal scores and contribute directly through $\phi_2(Z^P_{ij},Z^P_{ik};\rho)$. For binary responses, only the threshold regions are observed, so their contributions are the probabilities obtained by integrating over the corresponding regions.
Combining these two contributions into a pairwise pseudo-log-likelihood for each item pair, we maximize
\begin{equation}\label{eq:dim-pair}
\hat{\rho}_{jk}
=
\underset{-1<\rho<1}{\arg\max}\left(\sum_{i\in P}\log\phi_2(Z^P_{ij},Z^P_{ik};\rho)+\sum_{u,v\in\{0,1\}}n_{uv,jk}\log\pi_{uv,jk}(\rho)\right).
\end{equation}
The first term represents probability observations, and the second represents binary observations. 
The continuous term has the same dependence on $\rho$ as the Gaussian-copula log density, up to marginal terms independent of $\rho$. 
With only probability observations, this estimator reduces to the Pearson correlation of standardized normal scores (including its boundary limit). 
With only binary observations, it reduces to the two-step tetrachoric estimator with estimated thresholds held fixed \citep{olsson1979}. 
%Eq.~\eqref{eq:dim-pair} adapts the common latent-correlation principle to the present source-by-response-format design; it is not an estimator taken unchanged from \citet{fan2014latent}.

Finally, after the construction of the nearest PSD correlation matrix with Eq.~\eqref{eq: PSD}, we eigendecompose $\widehat R=V\Lambda V^\top$ and inspect the ordered eigenvalues $\lambda_1\geq\cdots\geq\lambda_J\geq0$ in a scree plot. Because the diagonal elements of $\widehat R$ equal one, $\sum_j\lambda_j=J$. 
We report the variance share of the first principal component, $\lambda_1/\sum_{j=1}^{J}\lambda_j=\lambda_1/J$, and the ratio of the two largest eigenvalues, $\lambda_1/\lambda_2$. Both summarize the dominance of the first principal component. 
Appendix~\ref{sec:mixed-dim-mmlu} additionally evaluates residuals from a separately fitted one-factor correlation model.

\subsection{Item correlations under a shared 2PL conditional mean}\label{sec:dim-simulation}

We examine whether a common 2PL conditional mean yields similar item-correlation spectra across the two response formats. 
The simulation uses 100 fixed standardized normal-quantile abilities and 100 items, with $a_j\sim U(0.5,1.5)$ and $b_j\sim N(0,0.65^2)$ clipped to $[-1.3,1.3]$. 
The binary-100 and probability-100 datasets were generated using the full bank of 100 respondents. 
For the mixed-100 condition, the respondent bank was divided into two groups of 50 according to ability: respondents were ordered by ability, and one respondent from each neighboring pair was assigned to each response source.
Binary responses were generated for the 50 respondents assigned to the binary source (binary-50), and probability responses were generated for the remaining 50 respondents (probability-50). 
The mixed-100 dataset combined these two subsets. 
The binary-50 and probability-50 conditions used the same respective subsets as in mixed-100. Across the 50 replicates, binary draws and the stratified source assignments were varied.

The probability response is the exact 2PL conditional mean, and the binary response is a Bernoulli draw with that mean:
\begin{equation}\label{eq:dim-exact}
 \begin{aligned}
 Q_{ij}&=\mu_j(\theta_i)
 =\operatorname{logit}^{-1}\!\{1.702a_j(\theta_i-b_j)\},\\
 Y_{ij}\mid\theta_i&\sim\operatorname{Bernoulli}\{\mu_j(\theta_i)\}.
 \end{aligned}
\end{equation}

Table~\ref{tab:dim-simulation} shows how the observation process shapes the spectrum with a single latent ability. 
With positive discriminations, the exact 2PL probabilities are strictly increasing functions of the same ability. 
Therefore, in the probability-only conditions, all items rank respondents identically. 
The resulting normal-score columns coincide, yielding a rank-one correlation matrix and $\lambda_1/J=1.000$.
On the other hand, in the binary-only conditions, item-specific Bernoulli variation attenuates the common rank pattern, and the estimated $\lambda_1/J$ is about 0.49 under the same one-ability model. 
In the balanced mixed-100 condition, the probability-side structure dominates the pooled estimate and $\lambda_1/J$ reaches 0.9872. 
These results demonstrate that, even under a common one-dimensional 2PL conditional-mean model, the estimated correlation structure and the dominance of the first principal component depend substantially on the response format. 
In particular, the probability observations yield a much more dominant first principal component than the binary observations do, and this pattern largely carries over to the mixed-format analysis.

\begin{table}[!htb]
\centering\small
\caption{Mean $\lambda_1/J$ over 50 replicates under the exact 2PL conditional mean in Eq.~\eqref{eq:dim-exact} with standard deviations on the same scale.}
\label{tab:dim-simulation}
\begin{tabular}{lrr}\toprule
Data format & Mean $\lambda_1/J$ & Standard deviation\\\midrule
Binary-100 & 0.4879 & 0.0117 \\
Probability-100 & 1.0000 & 0.0000 \\
Binary-50 & 0.4953 & 0.0137 \\
Probability-50 & 1.0000 & 0.0000 \\
Mixed-100 & 0.9872 & 0.0005 \\
\bottomrule

\end{tabular}
\end{table}

\begin{figure}[!htb]
\centering
\includegraphics[width=\linewidth]{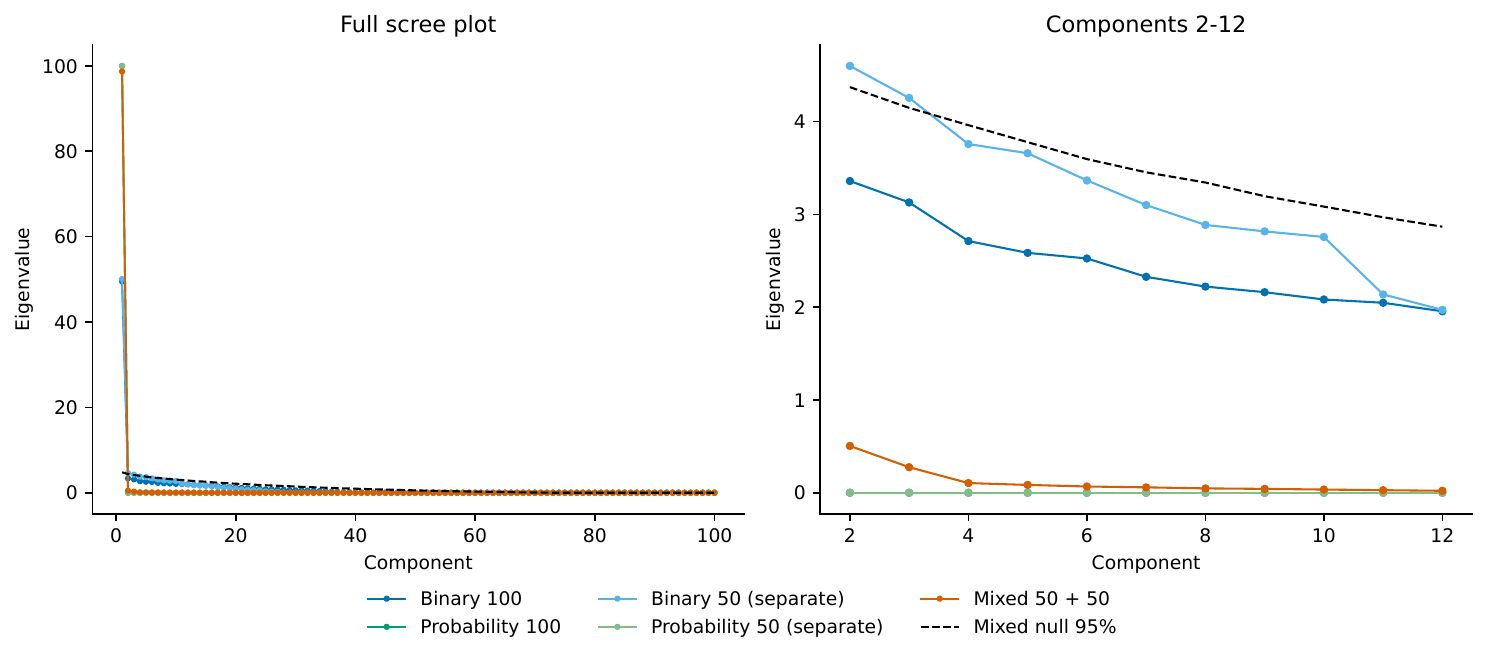}
\caption{Representative fixed-bank replicate under Eq.~\eqref{eq:dim-exact}: the full scree plot and components 2--12. 
Probability-50 and probability-100 overlap, with $\lambda_1/J=1.000$. 
The dashed null curve is the componentwise 95th percentile from 100 within-source column permutations for mixed-100, with the same correlation estimation and PSD correction applied to each permutation.
The reference applies to the illustrated replicate.}
\label{fig:dim-simulation}
\end{figure}

Figure~\ref{fig:dim-simulation} illustrates the spectrum in one replicate. 
Only the first eigenvalue exceeds the mixed-100 parallel-analysis reference, which breaks item associations by permuting responses within each source. 
This comparison describes the dominance of the first principal component in the illustrated replicate. 
%The simulation does not evaluate the power to detect a single latent dimension.

%Adding mean-preserving beta noise to the probability responses reduced the mixed-format $\lambda_1/J$ while retaining the same conditional mean and the same single ability. 
%Together, these results show why the magnitude of the $\lambda_1/J$ depends on the response-generating process. 
%Lower shares can arise under a one-factor model, and a dominant first principal component in the pooled matrix can reflect the way probabilities are observed.
\begin{comment}
\subsection{Implications for the MMLU application}

The MMLU analysis treats a common latent item correlation across Open and HELM as a working assumption. It remains unverified for these evaluation protocols. A pooled spectrum close to the Open-only spectrum can result from the much larger Open sample, even if the underlying item associations differ across sources.

Appendix~\ref{sec:mixed-dim-mmlu} compares the source-specific and pooled spectra on common items and examines residuals from a separate one-factor correlation fit. The spectra describe how strongly the first principal component dominates; the residuals describe how closely a one-factor model approximates the pairwise correlations. Both are exploratory diagnostics for assessing the plausibility of the IRT assumption.
\end{comment}

\section{MMLU dimensionality diagnostics across all 57 subjects}\label{sec:mixed-dim-mmlu}
For each of the 57 subjects, the source comparisons describe how the dominance of the first principal component varies across analyses, and the one-factor residuals summarize the fit to pairwise correlations on the retained items. 
%The common latent-correlation assumption remains unverified, so the two summaries provide exploratory evidence about dimensionality.

%\subsection{Common items, sample sizes, and source comparisons}
We applied Eq.~\eqref{eq:dim-pair} separately to HELM, Open, and their combination. 
For each MMLU subject, we matched items across the Open and HELM datasets and retained only items with nonconstant responses in both sources. 
This yielded 13,382 analyzable items out of the original 14,042, ensuring that the probability-only, binary-only, and integrated correlation analyses were compared on the same item sets. 
The analyses used approximately 6,480 Open models and 79 HELM models per subject.
For a diagnostic distinct from PCA, we also fit a unit-diagonal one-factor correlation model to the \emph{uncorrected} pairwise matrix $\widetilde{R}$~(not the PSD correlation matrix $\widehat{R}$) and calculate the standardized root mean square residual~(SRMR) as follows:
\begin{equation}
 \widehat{\boldsymbol l}=\arg\min_{|l_j|<1}\sum_{j<k}(\widetilde R_{jk}-l_jl_k)^2,
 \qquad
 \mathrm{SRMR}=\left\{\frac{2\sum_{j<k}(\widetilde R_{jk}-\widehat l_j\widehat l_k)^2}{J(J+1)}\right\}^{1/2}.
\end{equation}
Unique variances are $1-l_j^2$. 
We use the uncorrected pairwise correlation matrix to avoid masking discrepancies through PSD correction. 
The computational materials record absolute-residual quantiles and signed factor loadings
\footnote{
Because the one-factor correlation model allows loadings of either sign, it should be noted that a low SRMR does not necessarily support the positive-discrimination assumption of the fitted IRT model.
}. 

Table~\ref{tab:dim-all57} summarizes the dimensionality diagnostics for all MMLU subjects.

\begin{comment}
Normal-theory maximum-likelihood reference calculations of the comparative fit index (CFI), Tucker--Lewis index (TLI), and root mean square error of approximation (RMSEA) are retained as sensitivity diagnostics, not valid mixed-format goodness-of-fit tests. Formal tests would need to account for estimated margins and thresholds, dependence among pairwise estimates, and overlapping or related models. Singular correlation matrices also require numerical regularization that can alter the indices without materially changing the spectrum. We therefore use descriptive residuals and source comparisons without certifying a single-factor model for any subject. The computational specification records these reference calculations and their regularization settings.
\end{comment}
\clearpage
\begingroup\fontsize{10}{9.6}\selectfont\renewcommand{\arraystretch}{.95}\setlength{\LTpre}{0pt}\setlength{\tabcolsep}{2.3pt}\setlength{\LTcapwidth}{\linewidth}
\begin{longtable}{p{.29\linewidth}r*{9}{r}}
\caption{All-subject dimensionality diagnostics. 
B: HELM binary responses; 
P: probability data from Open LLM Leaderboard; 
I: integration of B and P. 
$\lambda_1/J$ and $\lambda_1/\lambda_2$ use PSD-corrected correlation matrix $\widehat{R}$; 
SRMR uses the uncorrected correlation matrix $\widetilde{R}$. 
}\label{tab:dim-all57}\\
\toprule Subject & $J$ & \multicolumn{3}{c}{$\lambda_1/J$} & \multicolumn{3}{c}{$\lambda_1/\lambda_2$} & \multicolumn{3}{c}{SRMR}\\
 & & B & P & I & B & P & I & B & P & I\\\midrule\endfirsthead
\multicolumn{11}{l}{\tablename\ \thetable\ (continued)}\\\toprule
Subject & $J$ & \multicolumn{3}{c}{$\lambda_1/J$} & \multicolumn{3}{c}{$\lambda_1/\lambda_2$} & \multicolumn{3}{c}{SRMR}\\
 & & B & P & I & B & P & I & B & P & I\\\midrule\endhead
\midrule\multicolumn{11}{r}{Continued on next page}\\\endfoot
\bottomrule\endlastfoot
Abstract algebra & 100 & 0.295 & 0.257 & 0.257 & 2.88 & 1.71 & 1.70 & 0.312 & 0.188 & 0.188 \\
Anatomy & 128 & 0.463 & 0.424 & 0.423 & 2.42 & 4.84 & 4.85 & 0.478 & 0.117 & 0.117 \\
Astronomy & 140 & 0.452 & 0.443 & 0.442 & 2.55 & 4.59 & 4.58 & 0.526 & 0.116 & 0.116 \\
Business ethics & 96 & 0.462 & 0.403 & 0.402 & 2.66 & 2.68 & 2.68 & 0.469 & 0.164 & 0.164 \\
Clinical knowledge & 252 & 0.481 & 0.470 & 0.470 & 2.43 & 6.09 & 6.09 & 0.485 & 0.105 & 0.105 \\
College biology & 139 & 0.518 & 0.556 & 0.555 & 2.68 & 7.74 & 7.72 & 0.466 & 0.090 & 0.090 \\
College chemistry & 95 & 0.287 & 0.340 & 0.340 & 1.38 & 3.97 & 3.97 & 0.507 & 0.127 & 0.127 \\
College computer science & 97 & 0.358 & 0.363 & 0.363 & 2.01 & 4.54 & 4.53 & 0.389 & 0.120 & 0.120 \\
College mathematics & 100 & 0.219 & 0.277 & 0.276 & 1.95 & 3.31 & 3.30 & 0.264 & 0.138 & 0.138 \\
College medicine & 164 & 0.444 & 0.487 & 0.487 & 2.13 & 7.12 & 7.11 & 0.484 & 0.094 & 0.094 \\
College physics & 102 & 0.318 & 0.313 & 0.312 & 2.39 & 3.07 & 3.06 & 0.424 & 0.147 & 0.147 \\
Computer security & 90 & 0.562 & 0.550 & 0.550 & 4.28 & 7.61 & 7.62 & 0.465 & 0.086 & 0.086 \\
Conceptual physics & 231 & 0.410 & 0.384 & 0.384 & 2.20 & 4.01 & 3.99 & 0.479 & 0.129 & 0.129 \\
Econometrics & 109 & 0.360 & 0.369 & 0.368 & 3.05 & 3.95 & 3.94 & 0.382 & 0.129 & 0.129 \\
Electrical engineering & 142 & 0.345 & 0.354 & 0.354 & 2.19 & 3.50 & 3.49 & 0.503 & 0.131 & 0.131 \\
Elementary mathematics & 376 & 0.367 & 0.317 & 0.317 & 3.09 & 2.85 & 2.84 & 0.323 & 0.149 & 0.149 \\
Formal logic & 125 & 0.320 & 0.320 & 0.319 & 2.51 & 3.33 & 3.32 & 0.398 & 0.139 & 0.139 \\
Global facts & 100 & 0.218 & 0.288 & 0.288 & 1.46 & 2.30 & 2.30 & 0.348 & 0.161 & 0.161 \\
High school biology & 285 & 0.503 & 0.530 & 0.530 & 2.44 & 7.40 & 7.39 & 0.491 & 0.092 & 0.092 \\
High school chemistry & 197 & 0.294 & 0.355 & 0.355 & 2.34 & 3.86 & 3.85 & 0.474 & 0.124 & 0.124 \\
High school computer science & 99 & 0.502 & 0.511 & 0.511 & 3.00 & 7.09 & 7.08 & 0.460 & 0.091 & 0.091 \\
High school european history & 157 & 0.500 & 0.574 & 0.574 & 1.68 & 8.90 & 8.91 & 0.510 & 0.084 & 0.084 \\
High school geography & 185 & 0.616 & 0.588 & 0.588 & 3.99 & 9.03 & 9.02 & 0.447 & 0.080 & 0.080 \\
High school government and politics & 159 & 0.644 & 0.627 & 0.627 & 3.47 & 10.78 & 10.78 & 0.486 & 0.075 & 0.075 \\
High school macroeconomics & 363 & 0.421 & 0.453 & 0.453 & 1.96 & 5.53 & 5.51 & 0.495 & 0.114 & 0.114 \\
High school mathematics & 270 & 0.210 & 0.211 & 0.210 & 1.64 & 1.88 & 1.87 & 0.274 & 0.172 & 0.172 \\
High school microeconomics & 226 & 0.483 & 0.456 & 0.456 & 2.58 & 4.36 & 4.35 & 0.473 & 0.124 & 0.124 \\
High school physics & 151 & 0.316 & 0.300 & 0.299 & 1.96 & 3.45 & 3.44 & 0.351 & 0.134 & 0.135 \\
High school psychology & 491 & 0.642 & 0.638 & 0.637 & 4.02 & 12.70 & 12.69 & 0.459 & 0.070 & 0.070 \\
High school statistics & 211 & 0.342 & 0.313 & 0.313 & 2.52 & 2.89 & 2.88 & 0.383 & 0.148 & 0.148 \\
High school US history & 201 & 0.527 & 0.596 & 0.596 & 1.73 & 11.02 & 11.02 & 0.520 & 0.077 & 0.077 \\
High school world history & 230 & 0.525 & 0.626 & 0.626 & 2.30 & 10.41 & 10.40 & 0.526 & 0.075 & 0.075 \\
Human aging & 212 & 0.427 & 0.543 & 0.542 & 2.37 & 8.62 & 8.61 & 0.536 & 0.088 & 0.088 \\
Human sexuality & 125 & 0.470 & 0.533 & 0.533 & 2.18 & 8.49 & 8.48 & 0.464 & 0.088 & 0.088 \\
International law & 115 & 0.479 & 0.568 & 0.568 & 3.16 & 6.84 & 6.85 & 0.553 & 0.093 & 0.093 \\
Jurisprudence & 102 & 0.576 & 0.518 & 0.518 & 2.69 & 7.33 & 7.33 & 0.485 & 0.091 & 0.091 \\
Logical fallacies & 149 & 0.482 & 0.542 & 0.542 & 2.52 & 9.17 & 9.17 & 0.518 & 0.084 & 0.084 \\
Machine learning & 112 & 0.331 & 0.346 & 0.346 & 1.89 & 3.38 & 3.38 & 0.452 & 0.148 & 0.148 \\
Management & 96 & 0.552 & 0.585 & 0.584 & 3.52 & 9.32 & 9.33 & 0.512 & 0.075 & 0.075 \\
Marketing & 209 & 0.675 & 0.627 & 0.627 & 4.27 & 12.77 & 12.78 & 0.456 & 0.068 & 0.068 \\
Medical genetics & 92 & 0.506 & 0.516 & 0.515 & 3.18 & 6.69 & 6.68 & 0.427 & 0.094 & 0.094 \\
Miscellaneous & 711 & 0.594 & 0.605 & 0.605 & 3.94 & 10.58 & 10.59 & 0.491 & 0.076 & 0.076 \\
Moral disputes & 331 & 0.431 & 0.517 & 0.517 & 2.60 & 8.47 & 8.46 & 0.521 & 0.094 & 0.094 \\
Moral scenarios & 895 & 0.393 & 0.356 & 0.355 & 2.33 & 2.58 & 2.57 & 0.320 & 0.197 & 0.197 \\
Nutrition & 290 & 0.400 & 0.446 & 0.446 & 1.96 & 5.87 & 5.87 & 0.548 & 0.105 & 0.105 \\
Philosophy & 296 & 0.435 & 0.491 & 0.491 & 2.75 & 7.08 & 7.08 & 0.532 & 0.097 & 0.097 \\
Prehistory & 298 & 0.498 & 0.490 & 0.490 & 2.90 & 6.12 & 6.11 & 0.528 & 0.104 & 0.104 \\
Professional accounting & 280 & 0.364 & 0.354 & 0.354 & 3.15 & 3.88 & 3.87 & 0.390 & 0.130 & 0.130 \\
Professional law & 1516 & 0.317 & 0.362 & 0.361 & 2.32 & 4.47 & 4.46 & 0.387 & 0.128 & 0.128 \\
Professional medicine & 261 & 0.356 & 0.453 & 0.453 & 1.51 & 4.83 & 4.82 & 0.504 & 0.119 & 0.120 \\
Professional psychology & 578 & 0.498 & 0.480 & 0.479 & 2.95 & 5.97 & 5.96 & 0.464 & 0.107 & 0.107 \\
Public relations & 102 & 0.412 & 0.481 & 0.481 & 2.56 & 5.97 & 5.97 & 0.551 & 0.099 & 0.099 \\
Security studies & 231 & 0.340 & 0.519 & 0.519 & 1.30 & 7.18 & 7.18 & 0.585 & 0.096 & 0.096 \\
Sociology & 187 & 0.466 & 0.604 & 0.604 & 2.32 & 11.40 & 11.42 & 0.604 & 0.076 & 0.076 \\
Us foreign policy & 91 & 0.647 & 0.641 & 0.641 & 6.02 & 11.02 & 11.03 & 0.424 & 0.071 & 0.071 \\
Virology & 135 & 0.344 & 0.509 & 0.509 & 2.18 & 8.17 & 8.18 & 0.639 & 0.085 & 0.085 \\
World religions & 157 & 0.567 & 0.598 & 0.598 & 4.91 & 11.75 & 11.77 & 0.493 & 0.071 & 0.071 \\

\end{longtable}\endgroup
\clearpage
\FloatBarrier

\end{document}